\documentclass[twocolumn,english,pra,aps,superscriptaddress]{revtex4-1}
\usepackage[T1]{fontenc}
\usepackage{color}
\usepackage{babel}
\usepackage{mathtools}
\usepackage{bm}
\usepackage{graphicx}
\usepackage{esint}
\usepackage{amsfonts}
\usepackage{graphicx}
\usepackage{float}
\usepackage{subfigure}
\usepackage{amsthm}
\usepackage{amssymb}

\usepackage[usenames,divipsnames,svgnames,table]{xcolor}
\usepackage[unicode=true,pdfusetitle,
bookmarks=true,bookmarksnumbered=false,bookmarksopen=false,
breaklinks=true,pdfborder={0 0 0},backref=false,colorlinks=true]{hyperref}
\hypersetup{linkcolor=NavyBlue,urlcolor=NavyBlue,citecolor=NavyBlue}
\usepackage{breakurl}

\newtheorem{theorem}{Theorem}
\newtheorem{corollary}{Corollary}

\begin{document}

\title{Bhatia-Davis formula in the quantum speed limit}

\author{Jing Liu}
\affiliation{MOE Key Laboratory of Fundamental Physical Quantities Measurement,
PGMF and School of Physics, Huazhong University of Science and Technology,
Wuhan 430074, China}

\author{Zibo Miao}
\affiliation{School of Mechanical Engineering and Automation, Harbin Institute
of Technology, Shenzhen, Shenzhen 518055, China}

\author{Libin Fu}
\affiliation{Graduate School, China Academy of Engineering Physics,
Beijing 100193, China}

\author{Xiaoguang Wang}
\affiliation{Zhejiang Institute of Modern Physics, Department of Physics,
Zhejiang University, Hangzhou 310027, China}

\begin{abstract}
The Bhatia-Davis theorem provides a useful upper bound for the variance in
mathematics, and in quantum mechanics, the variance of a Hamiltonian is naturally
connected to the quantum speed limit due to the Mandelstam-Tamm bound. Inspired
by this connection, we construct a formula, referred to as the Bhatia-Davis formula,
for the characterization of the quantum speed limit in the Bloch representation. We
first prove that the Bhatia-Davis formula is an upper bound for a recently proposed
operational definition of the quantum speed limit, which means it can be used to
reveal the closeness between the timescale of certain chosen states to the systematic
minimum timescale. In the case of the largest target angle, the Bhatia-Davis formula
is proved to be a valid lower bound for the evolution time to reach the target
when the energy structure is symmetric. Regarding few-level systems, it is
also proved to be a valid lower bound for any state in two-level systems
with any target, and for most mixed states with large target angles in equally
spaced three-level systems.
\end{abstract}

\maketitle

\section{Introduction}

How fast a quantum system can evolve as required is usually referred to as the
problem of quantum speed limit in quantum foundations. This problem is important
in quantum foundations as it is naturally related to the uncertainty relations
and other fundamental properties of quantum mechanics. For instance, the first
bound concerning quantum speed limit given by Mandelstam and Tamm in 1945~\cite{Mandelstam1945}
was derived based on the uncertainty relations. Nowadays, the quantum
speed limit has gone way beyond the quantum foundations and attracted much
attention from the community of quantum information and quantum technology due
to the fact that the existence of noise is the major obstacle in most quantum
information processing to provide true quantum advantages in practice, and fast
evolutions could be a very useful approach to reduce the effect of noise in these
processes and help to reveal the quantum advantage.

The historical development of quantum speed limit is basically the development
of mathematical tools. Various tools have been developed for different
scenarios~\cite{Deffner2017}, including unitary dynamics~\cite{Mandelstam1945,
Margolus1998,Giovannetti2004}, open systems~\cite{Taddei2013,Campo2013,Deffner2013,
Sun2015,Marvian2015,Mirkin2016,Campo2019,Campaioli2018,Campaioli2019,Chenu17,Beau17b,
Cai2017,Villamizar2015,Sun2019,Meng2015,Mirkin2020,Zhang2014,Liu2015,Wu2018,Mondal2016},
quantum metrology~\cite{Giovannetti2003,Giovannetti2006,Beau17a}, quantum
control~\cite{Caneva2009,Hegerfeldt2013,Funo17,Campbell2017,Poggi2019}, quantum
phase transitions~\cite{Heyl2017,Shao2020}, quantum information processing~\cite{Epstein2017,
Girolami2019,Ashhab2012}, quantum resources~\cite{Marvian2016}, geometry of
quantum mechanics~\cite{Pires2016,Bukov2019,Sun2021}, and even the classical
systems~\cite{Margolus11,Shanahan2018,Okuyama2018,Amari16}. The crossover between
quantum speed limits has been observed in an optically trapped single atom system
recently~\cite{Ness2021}. Most existing tools in this field can be divided into
the Mandelstam-Tamm type and Margolus-Levitin type, which originate from the
Mandelstam-Tamm bound $\pi/(2\Delta H)$~\cite{Mandelstam1945} and the Margolus-Levitin
bound $\pi/(2\langle H\rangle)$~\cite{Margolus1998}, where $\langle H\rangle$ is
the expected value of the Hamiltonian $H$ and $\Delta H:=\sqrt{\langle H^2\rangle
-\langle H\rangle^2}$ is the corresponding deviation. The major difference between
these two types is that the former one is depicted by the deviation and the latter
one uses only the expected value.

Different with these two types, an operational definition of quantum speed limit
was proposed in 2020~\cite{Shao2020} based on the optimization of states that
can fulfill the target. One advantage of this operational definition is its
independence of the quantum states, which means it is the systematic minimum
timescale for this target and determined only by the Hamiltonian structure.
However, in quantum technology, some specific quantum states, like NOON states,
cat states, or certain types of entangled states, may be more worth studying
than a general one in some scenarios, and it is also possible that one does
not care the systematic minimum timescale, but is more interested in the time
scale of these specific states. In such cases, state-dependent tools could be
more handy than the operational definition. In the meantime, since the operational
definition includes only the information of systematic minimum timescale and
corresponding optimal states, it cannot reflect the closeness of the timescales
between concerned states and the optimal ones. Hence, the state-dependent
tools, especially those can naturally connect to the operational definition,
would be very helpful to reveal this closeness and thus more useful in practice.
Searching such state-dependent tools is a major motivation of this paper.

In mathematics, for a set of bounded real numbers $\{x_i\}$ with the expected
value $\bar{x}=\frac{1}{n}\sum_{i}x_i$ ($n$ is the number of elements),
Bhatia and Davis provided a very useful upper bound on the variance
$\mathrm{var}(x)=\frac{1}{n}\sum_{i}(x_i-\bar{x})^2$ in 2000~\cite{Bhatia2000},
\begin{equation}
\mathrm{var}(x)\leq (M-\bar{x})(\bar{x}-m),
\end{equation}
where $m$, $M$ are the lower and upper bounds of the set, $m\leq x_i\leq M$
for any element $x_i$ in the set. This bound can be naturally extended to the
statistics and further to the quantum mechanics. In quantum mechanics, the
Bhatia-Davis inequality can be rewritten into $\Delta^2 H\leq(E_{\max}-\langle H\rangle)
(\langle H\rangle-E_{\min})$ with $E_{\max}$ and $E_{\min}$ the maximum and
minimum energies with respect to $H$. Compared to the Mandelstam-Tamm bound,
it is obvious that
\begin{equation}
\frac{\pi}{2\Delta H}\geq\frac{\pi}{2\sqrt{(E_{\max}-\langle H\rangle)
(\langle H\rangle-E_{\min})}};
\end{equation}
namely, the Bhatia-Davis inequality provides a lower bound for the Mandelstam-Tamm
bound. Physically, the Bhatia-Davis bound indicates that the evolution time for
a state to reach one of its orthogonal states is bounded by the gap between the
maximum (minimum) energy and the average energy. However, since the Mandelstam-Tamm
bound itself is attainable only for two-level pure states~\cite{Deffner2017,Levitin2009},
the Bhatia-Davis bound above would be more difficult to saturate, and thus lack
of practicability. Nevertheless, things are more complicated in the Bloch
representation, which gives us a chance to introduce a similar formula by
replacing $\pi$ to a general target angle $\Theta$ defined in the Bloch
representation, and thoroughly study its role in quantum speed limit. In the
entire paper this formula will be referred to as the Bhatia-Davis formula. The
connection between this formula and the operational definition will be studied,
along with its behaviors and roles in both multilevel and few-level systems from
the aspect of quantum speed limit.

\section{Bhatia-Davis formula}

\subsection{Upper bound of the OQSL}

The Bloch representation is a common geometric approach for quantum states,
and widely applied in many topics in quantum information. In this
representation, a $N$-dimensional density matrix $\rho$ can be expressed by
\begin{equation}
\rho=\frac{1}{N}\left(\openone+\sqrt{\frac{N(N-1)}{2}}\vec{r}\cdot\vec{\lambda}\right),
\end{equation}
where $\vec{r}$ is a real vector referred to as the Bloch vector, $\openone$ is
the identity matrix and $\vec{\lambda}$ is the $(N^2-1)$-dimensional vector of
SU(N) generators. Throughout this paper, the target angle is defined
by the angle between the Bloch vectors of the initial state $\vec{r}$ and its
evolved state $\vec{r}(t)$~\cite{Campaioli2018,Campaioli2019,Shao2020}
\begin{equation}
\theta(t,\vec{r})=\arccos\left(\frac{\vec{r}\cdot\vec{r}(t)}{|\vec{r}|
|\vec{r}(t)|}\right),
\label{eq:Bloch_angle}
\end{equation}
where $\theta(t, \vec{r})\in [0, \pi]$.

Inspired by the Bhatia-Davis inequality, we define the general form of Bhatia-Davis
formula in the Bloch representation as
\begin{equation}
\tau_{\mathrm{BD}}:=\frac{\Theta}{2\sqrt{(E_{\mathrm{max}}-\langle H\rangle)
(\langle H\rangle-E_{\min})}},
\label{eq:tauBD}
\end{equation}
where $E_{\mathrm{max}}$ and $E_{\min}$ are the highest and lowest energies of
the Hamiltonian $H$ and $\langle H\rangle=\mathrm{Tr}(\rho H)$ is the expected
value with respect to the state $\rho$. $\Theta$ is a fixed target angle. In the
entire paper, we denote $E_k$ ($|E_k\rangle$) as the $k$th eigenvalue (eigenstate)
of $H$ with $k\in[0,N-1]$. Without loss of generality, we assume $E_k\leq E_j$
for $k<j$ and there exist at least two different energy values in $H$, namely,
not all the equalities can be achieved simultaneously. Recently, Becker \emph{et
al.}~\cite{Becker2021} provided a quantum speed limit, which also contains the
maximum energy, with energy-constrained diamond norms between two unitaries.
Traditionally, whether a bound contains the variance or the expected value is a
major criterion to distinguish Mandelstam-Tamm type bounds and Margolus-Levitin
type bounds. From this perspective, the Bhatia-Davis formula looks more like a
Margolus-Levitin type as it contains the expected value. However, its dependence
on the maximum and minimum energies makes it not a typical one. Hence, it may be
more appropriate to be treated as a totally different type.

\begin{figure*}[tp]
\includegraphics[width=14cm]{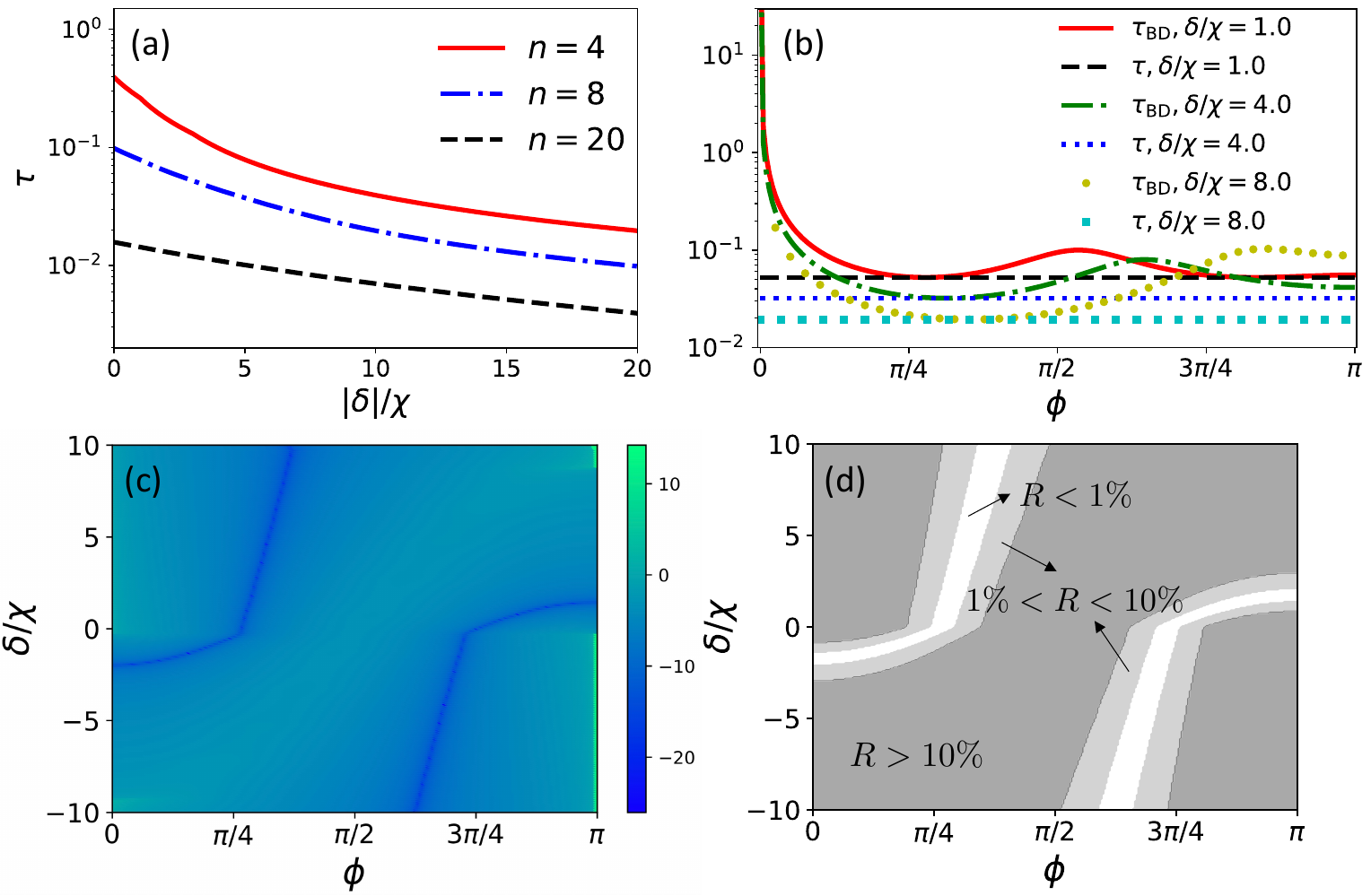}
\caption{(a) The OQSL $\tau$ as a function of $|\delta|/\chi$ for particle number
$n=4$ (solid red line), $n=8$ (dash-dotted blue line) and $n=20$ (dashed black
line). (b) The Bhatia-Davis formula $\tau_{\mathrm{BD}}$ and the OQSL $\tau$
as a function of $\phi$ for different values of $\delta/\chi$. The solid red, dash-dotted
green and circled yellow lines represent $\tau_{\mathrm{BD}}$ for $\delta/\chi=1.0,4.0,8.0$,
and the dashed black, dotted blue and squared cyan lines represent $\tau$ for
$\delta/\chi=1.0,4.0,8.0$, respectively. (c) The difference between $\tau_{\mathrm{BD}}$
and $\tau$ (in the scale of log) as a function of $\phi$ and $\delta/\chi$ for $n=10$.
(d) The regimes in (c) that the relative difference $R=(\tau_{\mathrm{BD}}-\tau)
/\tau<1\%$ (white area), $1\%<R<10\%$ (lightgray area) and $R>10\%$ (darkgray area).
$\chi$ is set to be 1 in all panels.
\label{fig:oat}}
\end{figure*}

Recently, an operational definition of the quantum speed limit (OQSL) was provided
and discussed~\cite{Shao2020}. The OQSL is defined via the set of states that can
reach the target angle $\mathcal{S}:=\{\vec{r}|\theta(t,\vec{r})=\Theta, \exists t\}$.
With this set, the OQSL (denoted by $\tau$) is defined as~\cite{Shao2020}
\begin{eqnarray}
\tau &=& \min_{\vec{r}\in\mathcal{S}} t \nonumber  \\
& & \mathrm{subject}~\mathrm{to}~~\theta(t,\vec{r})=\Theta.
\end{eqnarray}
The Bhatia-Davis formula has a natural connection with the OQSL due to the
following theorem.

\begin{theorem} \label{theorem:upperbound}
For time-independent Hamiltonians under unitary evolution, the Bhatia-Davis
formula is an upper bound for the OQSL:
\begin{equation}
\tau_{\mathrm{BD}} \geq \tau.
\end{equation}
\end{theorem}

The proof is given in Appendix~\ref{sec:apx_tauBD_tau}. The equality can be
attained when the average energy is half of the summation between the maximum
and minimum energies. This theorem indicates that $\tau_{\mathrm{BD}}$ can
reflect the closeness between the timescale of some specific states and the
systematic minimum timescale when the equality is attainable.

In the following we take the generalized one-axis twisting
model~\cite{Kitagawa1993,Ma2011,Jin2009} as an example to discuss this theorem.
The Hamiltonian for this model is
\begin{equation}
H=\chi J^2_z+\delta J_z,
\end{equation}
where $J_z=\sum^{n}_{i=1}\sigma^{(i)}_z/2$ with $\sigma^{(i)}_z$ the
Pauli matrix along $z$-direction for the $i$th spin and $\chi$, $\delta$ are the
coefficients. The Dicke state $|J=n/2,m\rangle$ ($m=0,\pm 1,\cdots,\pm J$ when
$n$ is even and $m=\pm 1/2, \pm 3/2,\cdots,\pm J$ when $n$ is odd) is the
eigenstate of $J_z$ with the corresponding eigenvalue $m$. For the sake of
simplicity, here we assume $n$ is even and $\chi>0$. According to Ref.~\cite{Shao2020},
the OQSL in this case can be expressed by $\tau=\Theta/(E_{\mathrm{max}}-E_{\min})$,
where the maximum energy reads
\begin{equation}
E_{\max}=\frac{1}{4}\left(\chi n^2+2|\delta| n\right),
\end{equation}
and the minimum energy reads
\begin{equation}
E_{\min}=\begin{cases}
\chi \mathcal{R}^2(\frac{\delta}{2\chi})-\delta\mathcal{R}(\frac{\delta}{2\chi}),
& \mathrm{for}~|\delta|/\chi \leq n, \\
\frac{1}{4}(\chi n^2-2|\delta|n), & \mathrm{for}~|\delta|/\chi > n.
\end{cases}
\end{equation}
Here $\mathcal{R}(\cdot)$ represents the function rounding to the nearest integer.
The OQSL can then be obtained correspondingly. When $\delta=0$, the Hamiltonian
is a standard one-axis twisting one, and the OQSL reduces to a simple form
\begin{equation}
\tau_{0}=\frac{4\Theta}{\chi n^2},
\end{equation}
which decreases quadratically with the growth of particle number $n$. As a
matter of fact, compared to $\tau_{0}$, the linear term $\delta J_z$ can
facilitate the reduction of the OQSL, as shown in Fig.~\ref{fig:oat}(a). For
example, in the case of a small $\delta$, $\tau\propto 1/(\chi n^2+2|\delta|n)$.
However, with the increase of $|\delta|$, when it is larger than $\chi n$, the
OQSL becomes
\begin{equation}
\tau_{1}=\frac{\Theta}{|\delta|n},
\end{equation}
which shows that the OQSL in this regime is not as sensitive as $\tau_0$ with
respect to $n$.

In this model, a well-used state is the coherent spin state $\exp(\zeta J_{+}
-\zeta^{*}J_{-})|J,J\rangle$ ($J_{\pm}=J_x\pm i J_y$). Since $\zeta$ can be
rewritten into $\zeta=-\frac{\phi}{2}\exp(-i\varphi)$ with $\phi\in[0,\pi]$
and $\varphi\in [0,2\pi]$, the coherent spin state can also be denoted by
$|\phi,\varphi\rangle$. For this state, the Bhatia-Davis formula
$\tau_{\mathrm{BD}}$ can be obtained by noticing the mean energy (details
are given in Appendix~\ref{sec:apx_tauBD_tau}) is
\begin{equation}
\langle H\rangle=\frac{1}{4}\left(2\delta n\cos\phi+\chi n^2\cos^2\phi
+\chi n\sin^2\phi\right),
\end{equation}
which indicates $\tau_{\mathrm{BD}}$ is not affected by $\varphi$.
Figure~\ref{fig:oat}(b) shows the values of $\tau_{\mathrm{BD}}$ and $\tau$ as
a function of $\phi$ for different values of $\delta$. In the case of $\delta=1.0$,
two regimes of $\phi$ around $\pi/4$ and $3\pi/4$ are optimal for $\tau_{\mathrm{BD}}$
to attain $\tau$. However, with the increase of $\delta$ ($4.0$ and $8.0$ in the plot),
the optimal regime around $\pi/4$ moves to the right and the optimal regime around
$3\pi/4$ vanishes completely. To provide a complete picture of the attainable
states, the difference between $\tau_{\mathrm{BD}}$ and $\tau$ (in the scale of
log) is given in Fig.~\ref{fig:oat}(c) as a function of $\phi$ and $\delta/\chi$,
which confirms that the attainable regime for a large $\phi$ vanishes with the
growth of $\delta$ when it is positive. As a matter of fact, most coherent spin
states have chances to be the attainable states when $\delta$ is tuned to proper
values. Particularly, the required optimal values of $\delta$ are very small when
$\phi$ is less than $\pi/4$ or larger than $3\pi/4$. The states around $\phi=\pi/2$
are more difficult to be the attainable states since they require large values
of $\delta$. However, although $\tau_{\mathrm{BD}}$ for these states are not
optimal, when $\delta/\chi$ is larger than, for example around $10.0$, there is
still a large regime around $\pi/2$ (on the left for $\delta/\chi>0$ and right
for $\delta/\chi<0$) in which the relative difference $R=(\tau_{\mathrm{BD}}-\tau)/\tau$
is less than $10\%$, as shown in Fig.~\ref{fig:oat}(d). In the meantime, the
area in the plot for the regime that $R<1\%$ is around $7.9\%$ ($R<10\%$ is
around $16.4\%$) of the total area, indicating that in this case, the timescale
for the coherent spin states to reach the target could be very close to the
systematic minimum time for a loose range of $\delta$. Another interesting
phenomenon is that the behavior of the difference between $\tau_{\mathrm{BD}}$
and $\tau$ is dramatically different for positive and negative signs of $\delta/\chi$
when it is not very large. $\tau_{\mathrm{BD}}$ is way closer to $\tau$ for a
negative (positive) $\delta/\chi$ when $\phi$ is small (large), which is due to
the fact that $\langle H\rangle$ is closer to $(E_{\max}+E_{\min})/2$ for a
negative (positive) value of $\delta/\chi$ when $\phi$ is small (large).

\subsection{Largest target angle $\Theta=\pi$}
\label{sec:multi}

In the study of quantum speed limit, the largest target angle $\Theta=\pi$
is worth paying particular attention as done in the Mandelstam-Tamm
bound~\cite{Mandelstam1945} and Margolus-Levitin bound~\cite{Margolus1998} since
it indicates the highest distinguishability. Due to the spirt of the operational
definition, the set of states that can fulfill the target. i.e., the set $\mathcal{S}$,
should be studied first as the state-dependent tools cannot reveal this information.
Considering the unitary evolution, we have the following observations on the set
$\mathcal{S}$ for the largest target $\Theta=\pi$.

\begin{theorem} \label{theorem:multiB}
For any finite-level Hamiltonian, there always exist states to fulfill the
target $\Theta=\pi$, i.e., the set $\mathcal{S}$ cannot be an empty set.
Furthermore, the set
\begin{equation*}
\mathcal{S}_0\!:=\!\{\vec{r}\,|r^2_{j^2+2k-1}\!+\!r^2_{j^2+2k}\!=\!|\vec{r}|^2,
\forall j\!\in\![1,\!N\!-\!1],k\!\in\![0,j\!-\!1]\}
\end{equation*}
is always a subset of $\mathcal{S}$:
\begin{equation}
\mathcal{S}_0\subseteq \mathcal{S}.
\end{equation}
\end{theorem}

Here $r_i$ is the $i$th entry of the Bloch vector $\vec{r}$. This theorem means
that any state in $\mathcal{S}_0$ can fulfill the target regardless of the
Hamiltonian structure. In the density matrix representation, the states in
$\mathcal{S}_0$ take the form
\begin{equation}
\left(\begin{array}{cccccc}
\frac{1}{N} & 0 & \cdots & \cdots & \cdots & 0 \\
0 & \frac{1}{N} & \vdots & \vdots & \vdots & \vdots\\
\vdots & \vdots & \ddots & \vdots & \rho_{kj} & \vdots\\
\vdots & \rho_{kj}^{*} & \vdots &\ddots  & \vdots & \vdots\\
\vdots & \vdots & \vdots & \vdots & \frac{1}{N}  & \vdots\\
0 & \cdots & \cdots & \cdots & \cdots & \frac{1}{N}
\end{array}\right)
\label{eq:state_S0}
\end{equation}
in the energy basis $\{|E_k\rangle\}$, where all the diagonal entries are $1/N$,
and the only nonzero nondiagonal entries are the $kj$th and $jk$th ones
with $j\in[1,N-1]$ and $k\in[0,j-1]$. Here $|\rho_{kj}|\in(0,1/N]$.

As a matter of fact, the theorem above also indicates that $\mathcal{S}_0$ is
the minimum set for $\mathcal{S}$, which leads to an interesting question that
what kind of Hamiltonians own the minimum set of $\mathcal{S}$? By denoting
$\mathcal{E}_{\mathrm{d}}$ as the set of all the values of energy differences,
\begin{equation}
\mathcal{E}_{\mathrm{d}}=\{E_j-E_k|\forall j\in[1,N-1],k\in[0,j-1]\},
\end{equation}
this question is answered by the corollary below.

\begin{corollary} \label{corollary:S0_1}
For the target $\Theta=\pi$, if a Hamiltonian satisfies that the ratio between
any two elements in $\mathcal{E}_{\mathrm{d}}$ cannot be written as the ratio
between two odd numbers,
\begin{equation}
\frac{E_{j_1}-E_{k_1}}{E_{j_2}-E_{k_2}}\neq \frac{2m_1+1}{2m_2+1}
\label{eq:Ed_cond}
\end{equation}
with $m_1,m_2$ any two non-negative integers for any two different groups of
subscripts $(j_1,k_1)$ and $(j_2,k_2)$, then
\begin{equation}
\mathcal{S}=\mathcal{S}_0.
\end{equation}
\end{corollary}

Here two different groups of subscripts means that $j_1=j_2$ and $k_1=k_2$
cannot hold simultaneously. The proofs of the theorem and corollary above
are given in Appendix~\ref{sec:apx_multilevel}. A natural Hamiltonian structure
to fit Eq.~(\ref{eq:Ed_cond}) is that all the elements in $\mathcal{E}_{\mathrm{d}}$
are noncommensurable to each other, which leads to the next corollary as follows.

\begin{corollary} \label{corollary:S0_2}
For the target $\Theta=\pi$ and the Hamiltonians with noncommensurable energy
differences, $\mathcal{S}=\mathcal{S}_0$.
\end{corollary}

Corollary~\ref{corollary:S0_1} could lead to the following no-go corollary for
multilevel systems (with at least three energy levels).

\begin{corollary} \label{corollary:nopure}
For multilevel systems with Hamiltonians stated in Corollary~\ref{corollary:S0_1},
no pure state can fulfill the target $\Theta=\pi$.
\end{corollary}

\begin{figure}[tp]
\centering
\includegraphics[width=8cm]{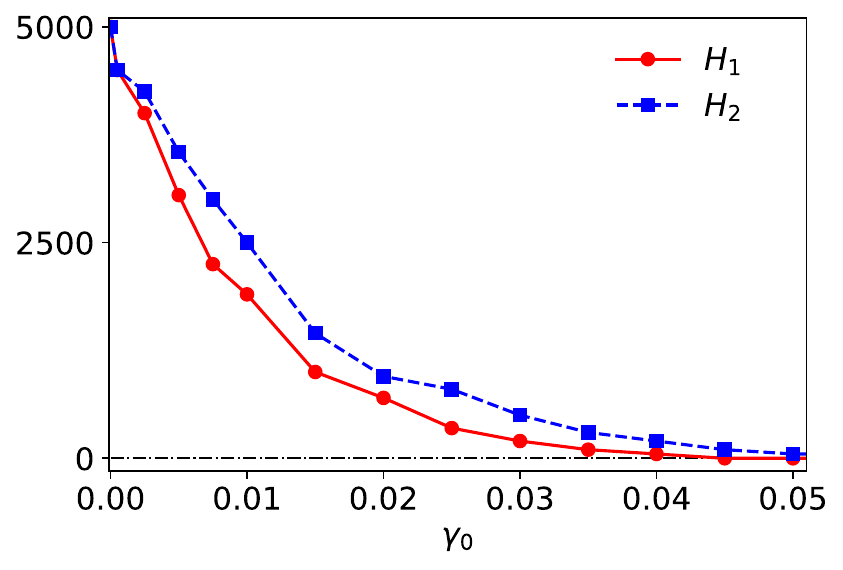}
\caption{The variety of state number capable of reaching
the target $\Theta=\pi$ with the change of decay rate $\gamma_0$
for the energy structures $\{1.0,2.1, 4.5, 8.3, 11.0\}$
($H_1$, red circles) and $\{1.0, 2\sqrt{7}, 6\sqrt{2},
6\sqrt{3}, 6\sqrt{5}\}$ ($H_2$, blue squares). $\bar{n}=1$ in
the plot.}
\label{fig:S_gamma}
\end{figure}

In practice, quantum systems are inevitably exposed to the environment and
therefore suffer from the noises. Hence, the performance of $\mathcal{S}$ must
be affected by the noise in general. The target $\Theta=\pi$ might be the most
sensitive case as it requires a large rotation of the Bloch vector, which may
not be possible in some type of noises. For example, if there exists a steady
state for some noisy dynamics, it is very possible that the states, whose angle
with the steady state are less than $\pi$, can never reach the target during
the evolution for a large enough decay rate. Hence, the state number in the
reachable state set $\mathcal{S}$ could be very limited, and even vanish in
such cases. For the sake of a more intuitive understanding, here we take the
damped five-level system as an example. The decoherence is described by the
master equation~\cite{Breuer2007}
\begin{align}
\partial_t\rho &=-i[H,\rho]+\gamma_0 (\bar{n}+1)\left(a\rho a^{\dagger}
-\frac{1}{2}a^{\dagger}a\rho-\frac{1}{2}\rho a^{\dagger}a\right) \nonumber \\
&+\gamma_0\bar{n}\left(a^{\dagger}\rho a-\frac{1}{2}aa^{\dagger}\rho
-\frac{1}{2}\rho aa^{\dagger}\right),
\end{align}
where $a$ ($a^{\dagger}$) is the lowering (raising) operator, $\gamma_0$ is a
constant decay rate, and $\bar{n}=1/[e^{\omega_0/(k_{\mathrm{B}}T)}-1]$ is
the Planck distribution with $k_{\mathrm{B}}$ the Boltzmann constant and $T$ the
temperature. Now consider two groups of energies $\{E_k\}=\{1.0,2.1,4.5,8.3,11.0\}$
(denoted by $H_1$) and $\{E_k\}=\{1.0,2\sqrt{7},6\sqrt{2},6\sqrt{3},6\sqrt{5}\}$
(denoted by $H_2$). According to Corollaries~\ref{corollary:S0_1} and~\ref{corollary:S0_2},
only the states in the form of Eq.~(\ref{eq:state_S0}), namely in the set $\mathcal{S}_0$,
can reach the target $\Theta=\pi$ in the unitary dynamics. To show the influence
of noise on $\mathcal{S}$, $5000$ random states in $\mathcal{S}_0$ are used to
test the attainability of the target $\Theta=\pi$, as given in Fig.~\ref{fig:S_gamma},
for $H_1$ (red circles) and $H_2$ (blue squares). $\bar{n}$ is set to be 1 in
the figure. In the absence of noise ($\gamma_0=0$), all states can reach
the target, just as Theorem~\ref{theorem:multiB} stated. With the increase of
decay rate, the number of states capable of reaching the target reduces in an
approximately exponential way. When $\gamma_0=0.05$, a very limited number of
states can still reach the target, and with the further increase of $\gamma_0$,
no state can ever reach the target eventually.

With the knowledge of $\mathcal{S}$, we could further study the Bhatia-Davis
formula. In general, the Bhatia-Davis formula is not a valid lower bound for
the evolution time to reach any target $\Theta$ due to Theorem~\ref{theorem:upperbound}.
For those Hamiltonians that the equality in Theorem~\ref{theorem:upperbound} is
not attainable, $\tau_{\mathrm{BD}}$ is always larger than $\tau$. In this case,
$\tau_{\mathrm{BD}}$ fails to be a valid lower bound for those states that
reaches the OQSL, and hence not a lower bound in general. However, for the
Hamiltonians and states that the equality can hold, $\tau_{\mathrm{BD}}$ might
still be a valid lower bound. One useful scenario is demonstrated as follows.

\begin{theorem} \label{theorem:tauBD_pi}
For a finite-level Hamiltonian that the energies are symmetric about
$\langle H\rangle$, the Bhatia-Davis formula is a valid lower bound for the
evolution time to reach the target $\Theta=\pi$, and for the states in
$\mathcal{S}$, it reduces to the OQSL.
\end{theorem}

The proof is given in Appendix~\ref{sec:apx_multilevel}. For such a symmetric
spaced energy structure, $\mathcal{S}$ must be larger than $\mathcal{S}_0$ since
$E_{N-1-k}-E_{N-1-j}=E_j-E_k$ for any $k<j\leq\left\lfloor\frac{N-1}{2}\right\rfloor$
with $\lfloor\cdot\rfloor$ the floor function. In this case, apart
from the states in Eq.~(\ref{eq:state_S0}), the states
\begin{equation*}
\left(\begin{array}{ccccccc}
\frac{1}{N} & 0 & \cdots & \cdots & \cdots & \cdots & 0\\
0 & \frac{1}{N} & \vdots & \vdots & \vdots & \vdots & \vdots\\
\vdots & \vdots & \ddots & \rho_{kj} & \vdots & \vdots & \vdots\\
\vdots & \vdots & \rho^*_{kj} & \ddots & \rho_{N-1-j,N-1-k} & \vdots & \vdots\\
\vdots & \vdots & \vdots & \rho^*_{N-1-j,N-1-k} & \ddots & \vdots & \vdots\\
\vdots & \vdots & \vdots & \vdots & \vdots  & \ddots & \vdots\\
0 & \cdots & \cdots & \cdots & \cdots & \cdots & \frac{1}{N}
\end{array}\right)
\end{equation*}
are also always capable of reaching the target. An typical form of this symmetric
structure is the equally spaced structure, i.e., $E_{k+1}-E_k$ is a constant for
any legitimate $k$. Hence, one can immediately obtain the following corollary.

\begin{corollary} \label{corollary:equal_spaced}
For an equally spaced finite-level Hamiltonian, the Bhatia-Davis formula is a
valid lower bound for the evolution time to reach the target $\Theta=\pi$, and
for the states in $\mathcal{S}$, it reduces to the OQSL.
\end{corollary}

\begin{figure}[tp]
\centering
\includegraphics[width=8.5cm]{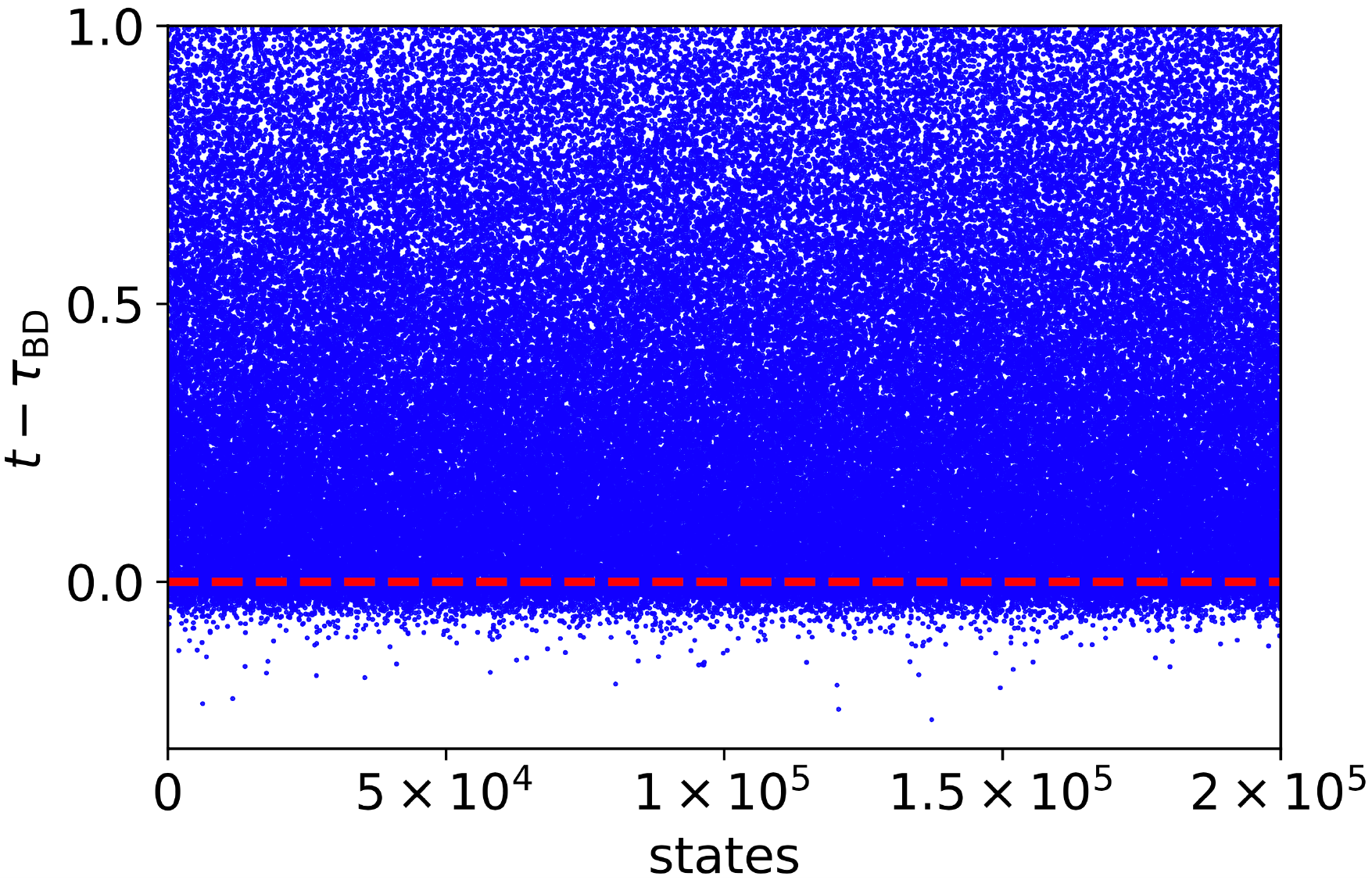}
\caption{The difference between the evolution time to reach
the target ($t$) and the Bhatia-Davis formula ($\tau_{\mathrm{BD}}$)
for $2\times 10^{5}$ pairs of randomly generated five-level
Hamiltonians and random states in $\mathcal{S}_0$. }
\label{fig:tauBD_test}
\end{figure}

Although the Bhatia-Davis formula is not a valid lower bound in general, how
general it is still keeps undiscovered. For this sake, we numerically test
whether $\tau_{\mathrm{BD}}$ is a valid lower bound for randomly generated
five-level Hamiltonians and random initial states with the target $\Theta=\pi$.
Since most randomly generated Hamiltonians satisfy the condition (\ref{eq:Ed_cond})
in Corollary~\ref{corollary:S0_1}, most random initial states cannot fulfill the
target expect for those in $\mathcal{S}_0$. This means $\tau_{\mathrm{BD}}$ is
indeed a formal lower bound for these states as the true evolution time to reach
the target is infinity. Hence, we pick only the random states in $\mathcal{S}_0$
for the test. Here $2\times 10^{5}$ random pairs of Hamiltonians and initial states
in $\mathcal{S}_0$ are generated and tested, as shown in Fig.~\ref{fig:tauBD_test}.
It can be seen that even limited in the set $\mathcal{S}_0$, the difference between
the evolution time $t$ (to reach the target) and $\tau_{\mathrm{BD}}$ is positive
for most states (around $90\%$), and therefore $\tau_{\mathrm{BD}}$ is a valid
lower bound for these $90\%$ states.

\section{Few-level systems}

\subsection{Two-level systems}
\label{sec:twolevel}

The two-level system is the most well-studied model in the topic of quantum
speed limit, and the only system that the well-known tools like the Mandelstam-Tamm
bound~\cite{Mandelstam1945} and the Margolus-Levitin bound~\cite{Margolus1998}
can be attained~\cite{Deffner2017,Levitin2009}. In this system, denoting the
ground and excited energies by $E_0$ and $E_1$ with the corresponding energy
levels $|E_0\rangle$ and $|E_1\rangle$, we then have the following theorem.

\begin{theorem}  \label{theorem:tauBD_2level}
For a two-level system under unitary evolution, the Bhatia-Davis formula is a
valid lower bound for the evolution time to reach any target angle $\Theta$,
and it can be attained by the states $(|E_0\rangle+e^{i\phi}|E_1\rangle)/\sqrt{2}$
with $\phi\in [0,2\pi)$.
\end{theorem}

The proof of this theorem is given in Appendix~\ref{sec:apx_twolevel}. The
attainability condition above comes from the requirement $\langle H\rangle
=(E_{0}+E_{1})/2$. A corollary with respect to OQSL can be immediately
obtained from this attainability condition.

\begin{corollary}
For a two-level system under unitary evolution, the Bhatia-Davis formula (bound)
reduces to OQSL when it is attainable.
\end{corollary}

With respect to the state-dependent bounds, several unified bounds have been
developed. For example, Sun \emph{et al.}~\cite{Sun2021} provide an unified bound
based on the changing rate of the phase in 2021. In the case of two-level systems,
a well-known unified tool for quantum speed limit is~\cite{Giovannetti2003,Giovannetti2004}
\begin{equation}
\tau_{\mathrm{C}}=\max\left\{\frac{\mathcal{A}}{\Delta H}, \frac{2\mathcal{A}^2}
{\pi\langle H\rangle}\right\},
\end{equation}
in which the target angle is defined via the Bures angle $\mathcal{A}=\arccos f$
with $f=\mathrm{Tr}\sqrt{\sqrt{\rho_0}\rho_1\sqrt{\rho_0}}$ the fidelity between
two quantum states $\rho_0$ and $\rho_1$. In the meantime, another tool based
on Bures angle is~\cite{Taddei2013}
\begin{equation}
\tau_{\mathrm{F}}=\frac{2\mathcal{A}}{\sqrt{F}},
\end{equation}
with $F$ the quantum Fisher information with respect to the evolution time $t$.
It is defined by $F=\mathrm{Tr}(\rho L^2)$ with $L$ the symmetric logarithmic
derivative. Here $L$ is determined by the equation $\partial_t \rho=(\rho L+L\rho)/2$.
In the Bloch sphere ($|E_1\rangle$ as the north pole), the density matrix is
$\rho=(\openone+\vec{r}\cdot\vec{\sigma})/2$ with
$\sigma=(\sigma_x, \sigma_y, \sigma_z)$ the vector of Pauli matrices. Then
$\tau_{\mathrm{F}}$ can be expressed by
\begin{equation}
\tau_{\mathrm{F}}=\frac{2\mathcal{A}}{(E_1-E_0)\sqrt{|\vec{r}|^2-r^2_z}},
\end{equation}
which is larger than $2\mathcal{A}/\Delta H$ since here $\Delta H=(E_1-E_0)
\sqrt{1-r^2_z}/2$. They are equivalent when the initial state is pure.
Combining several tools to construct a tighter bound for the quantum speed
limit is a common method in the previous studies in this field. Using this
strategy, the quantity
\begin{equation}
\tau_{\mathrm{m}}=\max\left\{\tau_{\mathrm{BD}},\tau_{\mathrm{F}},
\frac{2\mathcal{A}^2(\Theta)}{\pi\langle H\rangle}\right\}
\end{equation}
is a valid lower bound for the evolution time to reach the target in the case
of two-level systems. Notice that $\tau_{\mathrm{BD}}$ is not be a valid
lower bound in general for multilevel systems as discussed in the
previous section, hence $\tau_{\mathrm{m}}$ cannot be directly extended to
multilevel systems. Here $\mathcal{A}(\Theta)$ means the target is still
defined via Eq.~(\ref{eq:Bloch_angle}) and the value of $\mathcal{A}$ is
calculated via $\Theta$ and the initial state. As a matter of fact, the fidelity
between two qubits can be expressed by $f^2=\mathrm{Tr}(\rho_0\rho_1)+2\sqrt{\det(\rho_0)
\det(\rho_1)}$ with $\det(\cdot)$ the determinant~\cite{Hubner1992,Hubner1993}.
For unitary evolutions, it can be rewritten with the Bloch vectors into
$f^2=1-\frac{1}{2}|\vec{r}|^2(1-\cos\theta)$, where $|\vec{r}|$ is the norm
of the initial state. Hence, $\mathcal{A}(\Theta)=\arccos\sqrt{1-\frac{1}{2}
|\vec{r}|^2(1-\cos\Theta)}$, which indicates $\mathcal{A}\leq \Theta/2$ and the
equality holds for pure states. Because of this property, the following corollary
holds.

\begin{figure}[tp]
\includegraphics[width=8cm]{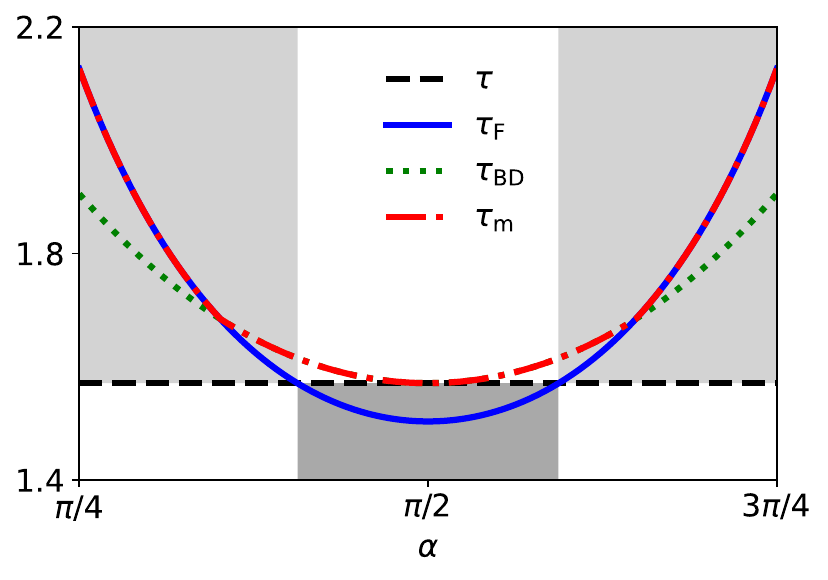}
\caption{The behaviors of different tools as a function of $\alpha$.
The dashed black, solid blue, dotted green and dash-dotted red lines
represent the OQSL $\tau$, the bound based on quantum Fisher information
$\tau_{\mathrm{F}}$, the Bhatia-Davis formula $\tau_{\mathrm{BD}}$, and
the combined formula $\tau_{\mathrm{m}}$. The parameters are set as
$E_1=2.0$, $E_0=1.0$, $|\vec{r}|=0.8$, and the target angle $\Theta=\pi/2$.
\label{fig:qubit}}
\end{figure}

\begin{corollary}
The Bhatia-Davis formula (bound) is equivalent to $\tau_{\mathrm{F}}$ for
two-level pure states.
\end{corollary}

With this corollary, and noticing that $\tau_{\mathrm{F}}=\mathcal{A}/\Delta H$
for two-level pure states, it is easy to see that $\tau_\mathrm{m}$ reduces to
$\tau_{\mathrm{C}}$ for two-level pure states. For mixed states, the relation
between $\tau_{\mathrm{BD}}$, $\tau_{\mathrm{F}}$ and $2\mathcal{A}^2/(\pi\langle H\rangle)$
is undetermined. However, in many cases, for instance when $(E_1+E_0)/(E_1-E_0)
>\sqrt{2}|\vec{r}|$ is satisfied, $\tau_{\mathrm{F}}$ is always larger than $2\mathcal{A}^2
/(\pi\langle H\rangle)$, and the value of $\tau_{\mathrm{m}}$ is taken as the
larger one between $\tau_{\mathrm{F}}$ and $\tau_{\mathrm{BD}}$. More calculation
details can be found in Appendix~\ref{sec:apx_twolevel}. An example is shown in
Fig.~\ref{fig:qubit} for the sake of a more intuitive understanding on $\tau_{\mathrm{m}}$.
Here $\alpha$ is the angle between the Bloch vector $|\vec{r}|$ and $z$-axis.
The states in the regime $\alpha\in[\Theta/2,\pi-\Theta/2]$ can fulfill
the target~\cite{Shao2020}. This plot first confirms that $\tau_{\mathrm{BD}}$
(dotted green line) is an upper bound for the OQSL $\tau$ (dashed black line).
However, $\tau_{\mathrm{F}}$ (solid blue line) and $\tau_{\mathrm{BD}}$ do
not have a fixed relation. $\tau_{\mathrm{F}}$ is less than $\tau_{\mathrm{BD}}$
for the states close to the $xy$ plane. Hence, $\tau_{\mathrm{m}}$ (dash-dotted
red line) equals $\tau_{\mathrm{BD}}$ in this regime, and it is indeed the
tightest bound for the evolution time. $2\mathcal{A}^2/(\pi\langle H\rangle)$ is
not plotted due to the fact that it is significantly smaller than the others in
this case.

Another advantage of using $\tau_{\mathrm{BD}}$ to construct the bound is that
$\tau_{\mathrm{m}}$ is always larger than $\tau$ due to Theorem 1, and it
reduces to $\tau$ in the $xy$ plane because of the attainability of
$\tau_{\mathrm{BD}}$, which means $\tau_{\mathrm{m}}$ can reflect the fact that
$\tau$ is the systematic minimum time to reach the target even when it is not
attainable. In the meantime, $\tau_{\mathrm{F}}$ can show this capability only
for some states (lightgray area), and it fails to do that for the states close
to the $xy$ plane (darkgray area) as it is smaller than $\tau$ in this regime.
Therefore, in the case where $\tau$ is not known, $\tau_{\mathrm{F}}$ cannot be
used to estimate the true minimum timescale.

\subsection{Three-level systems}

\begin{figure}[tp]
\includegraphics[width=8.5cm]{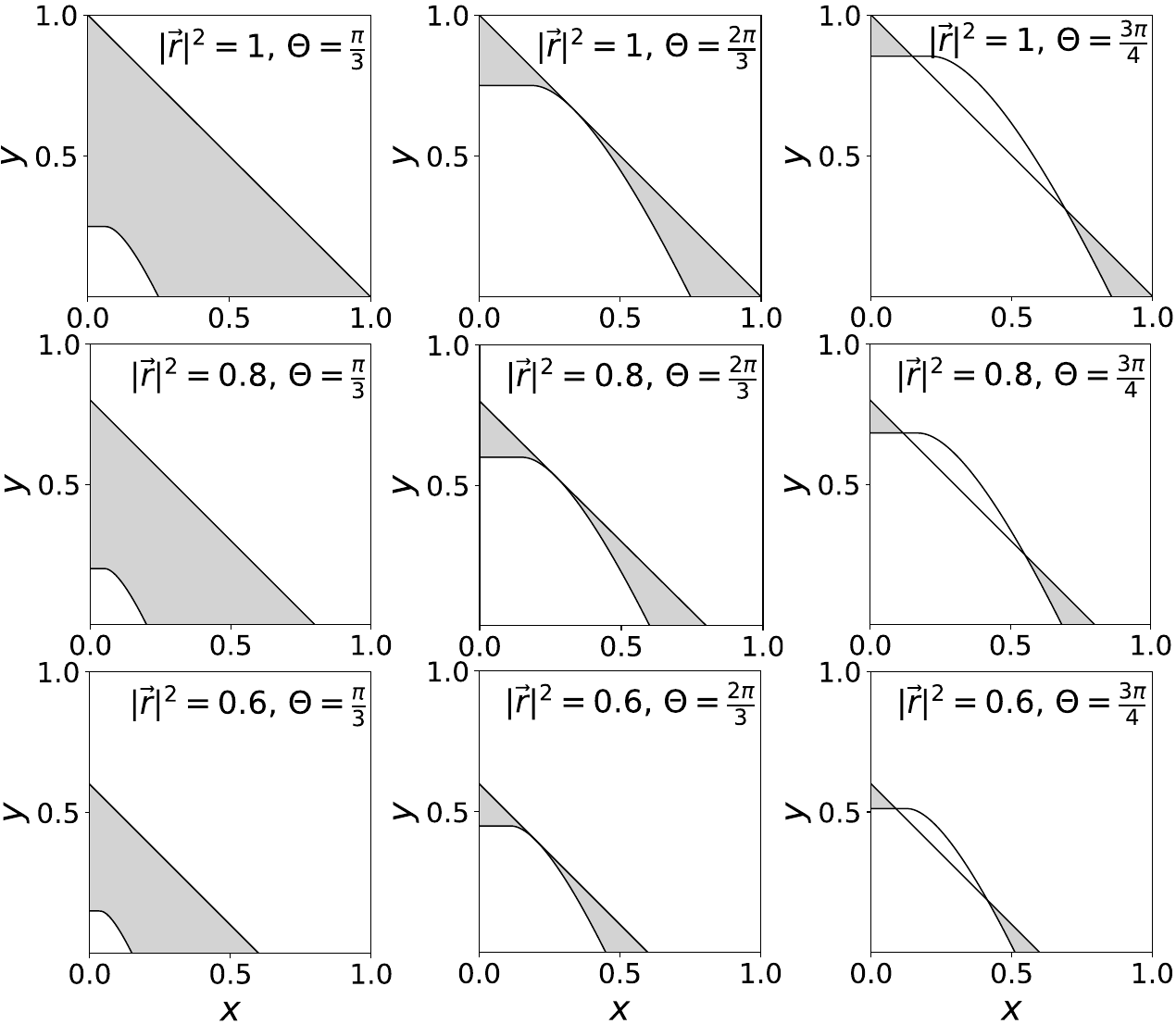}
\caption{Regimes of $x$, $y$ (gray areas) in the reachable states set
$\mathcal{S}$ for equally spaced three-level Hamiltonians with different
values of $|\vec{r}|^2$ and $\Theta$. $x=r^2_3+r^2_4$ and $y=r^2_0+r^2_1+r^2_5+r^2_6$
in the plot.}
\label{fig:S_3level}
\end{figure}

Three-level systems are very common in the study of quantum optics and quantum
information. As is the same as the general case, the Bhatia-Davis formula
\begin{equation}
\tau_{\mathrm{BD}}=\frac{\Theta}{2\sqrt{(E_2-\langle H\rangle)
(\langle H\rangle-E_{0})}}
\end{equation}
is not a valid lower bound in a general three-level system. However, Corollary~\ref{corollary:equal_spaced}
shows that $\tau_{\mathrm{BD}}$ is indeed a valid lower bound in the equally
spaced three-level systems for $\Theta=\pi$. To find out if $\tau_{\mathrm{BD}}$
still remains a valid bound for a general target in this case, we need to
study the set $\mathcal{S}$ first. Define $x=r^2_3+r^2_4$ and $y=r^2_0+r^2_1+r^2_5+r^2_6$
with $r_i$ ($i=0,\dots,6$) an entry of the Bloch vector, then for the states in
$\mathcal{S}$, $x$ and $y$ must locate in the following two regimes
\begin{equation}
\begin{cases}
y \geq 4|\vec{r}|\sin\left(\frac{\Theta}{2}\right)\sqrt{x}-4x, \\
y\leq 4x, \\
y\leq |\vec{r}|^2-x,
\end{cases} \label{eq:regime1}
\end{equation}
and
\begin{equation}
\begin{cases}
y \geq 4|\vec{r}|\sin\left(\frac{\Theta}{2}\right)\sqrt{x}-4x, \\
y\geq 4x, \\
y\geq |\vec{r}|^2\sin^2\left(\frac{\Theta}{2}\right), \\
y\leq |\vec{r}|^2-x.
\end{cases} \label{eq:regime2}
\end{equation}
The thorough derivation is given in Appendix~\ref{sec:apx_3level}. The full
regime is illustrated in Fig.~\ref{fig:S_3level} (gray areas) for different
values of $|\vec{r}|^2$ and $\Theta$. For the same $\Theta$ (columns in the plot),
the shapes of the regime basically coincide for different values of $|\vec{r}|^2$,
and its area shrinks with the reduction of $|\vec{r}|^2$. On the other hand, for
the same $|\vec{r}|^2$ (rows in the plot), the regime becomes narrower when
the value of $\Theta$ increases. These behaviors can also be reflected by the
variety of the ranges of $x$ ($y$) along the $x$ axis ($y$ axis), which is in
the regime $[|\vec{r}|^2\sin^2(\frac{\Theta}{2}),|\vec{r}|^2]$. The target $\Theta$
affects only the lower bound of this regime, which increases with the growth of
$\Theta$. Hence, the area of the full regime becomes smaller when $\Theta$ gets
larger. In the meantime, both bounds of this regime are affected by $|\vec{r}|^2$,
and the largest range is attained at $|\vec{r}|^2=1$, indicating that there exist
more choices of $x$, $y$ for pure states to reach the target.

Another interesting fact is that the full regime is continuous for the targets
less than $2\pi/3$, and it splits into two areas for those larger than $2\pi/3$.
This phenomenon is due to the fact that the minimum difference between $|\vec{r}|^2-x$
and $4|\vec{r}|\sin\left(\frac{\Theta}{2}\right)\sqrt{x}-4x$ with respect to $x$
is $|\vec{r}|^2\left[1-\frac{4}{3}\sin^2\left(\frac{\Theta}{2}\right)\right]$.
When $\Theta\leq 2\pi/3$, this minimum value is always positive, indicating
that all points on the line $y=4|\vec{r}|\sin\left(\frac{\Theta}{2}\right)\sqrt{x}-4x$
are feasible points in $\mathcal{S}$. By contrast, when $\Theta> 2\pi/3$,
only some points on this line are feasible, and the full regime then splits into
two areas.

With the information of $\mathcal{S}$, we then provide the following theorem on
the Bhatia-Davis formula.

\begin{theorem} \label{lamma_threelevel}
For an equally spaced three-level system with a gap $\Delta$, the Bhatia-Davis
formula is upper bounded by $\pi/(2\Delta)$,
\begin{equation}
\tau_{\mathrm{BD}}\leq\frac{\pi}{2\Delta},
\end{equation}
for any $|\vec{r}|\in(0,1]$ and $\Theta\in(0,\pi]$.
\end{theorem}

The derivation is given in Appendix~\ref{sec:apx_3level}. Since $\tau$ is upper
bouned by $\tau_{\mathrm{BD}}$ according to Theorem~\ref{theorem:upperbound}, this
result directly leads to $\tau\leq\pi/(2\Delta)$, which is fully reasonable
as $\tau=\Theta/(2\Delta)$ in this case~\cite{Shao2020}. Next we provide a theorem
to show when $\tau_{\mathrm{BD}}$ is a valid lower bound for a general target.

\begin{figure}[tp]
\includegraphics[width=8cm]{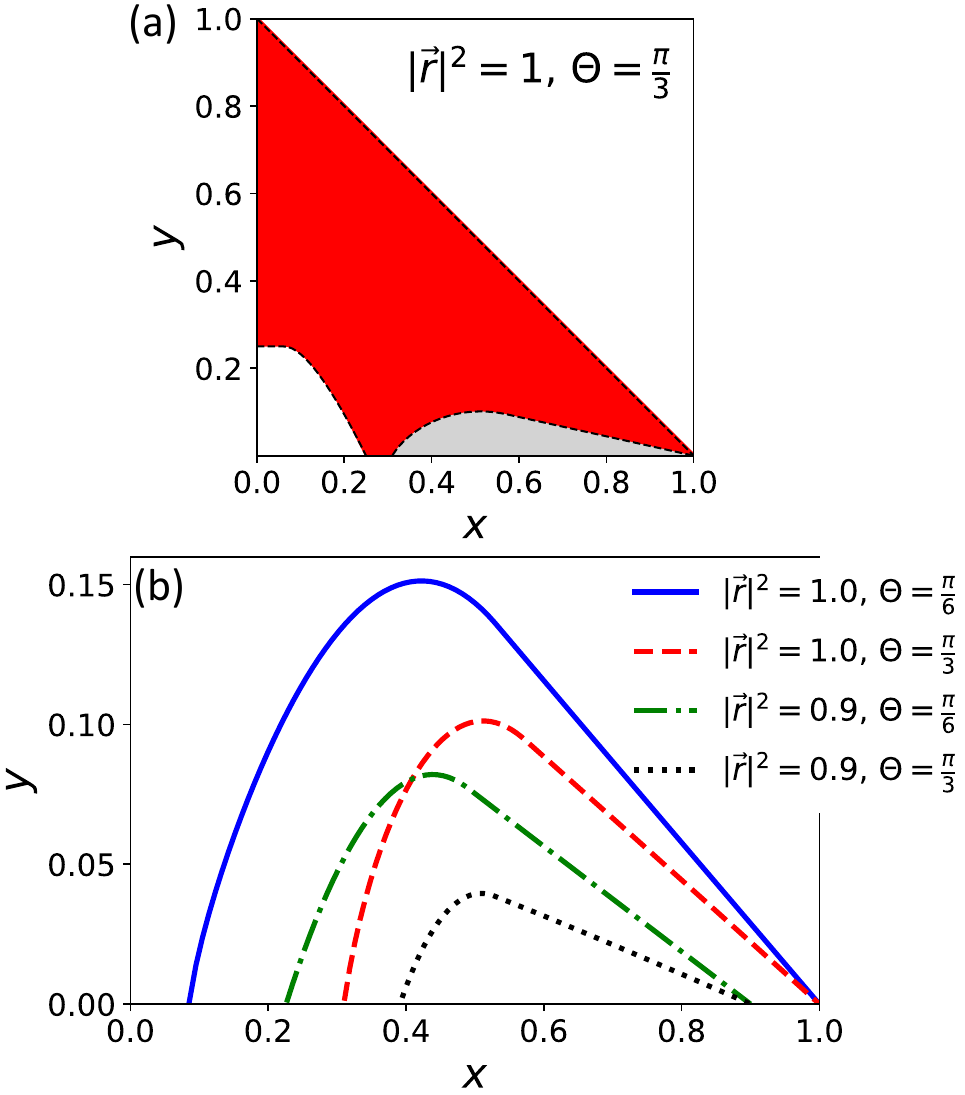}
\caption{(a) Regimes of $x$, $y$ (red/dark gray areas in online/print version)
that $\tau_{\mathrm{BD}}$ is always a valid lower bound in the case that
$|\vec{r}|^2=1$ and $\Theta=\pi/3$. $x=r^2_3+r^2_4$ and $y=r^2_0+r^2_1+r^2_5+r^2_6$
in the plot. (b) The variety of borderline given in Eq.~(\ref{eq:borderline})
as a function of $|\vec{r}|^2$ and $\Theta$. }
\label{fig:3level_border}
\end{figure}

\begin{theorem} \label{theorem:3level}
For an equally spaced three-level system with the gap $\Delta$, the Bhatia-Davis
formula is a valid lower bound for the evolution time to reach any target
$\Theta\in(0,\pi]$ at least for the states in the regimes
\begin{equation}
\begin{cases}
y \geq 4|\vec{r}|\sin\left(\frac{\Theta}{2}\right)\sqrt{x}-4x, \\
y\geq 4x, \\
y\geq |\vec{r}|^2\sin^2\left(\frac{\Theta}{2}\right), \\
y\leq |\vec{r}|^2-x,
\end{cases}
\end{equation}
and
\begin{equation}
\begin{cases}
y \geq 4|\vec{r}|\sin\left(\frac{\Theta}{2}\right)\sqrt{x}-4x, \\
y\leq 4x, \\
y\leq |\vec{r}|^2-x, \\
y\sin^2(\frac{\Delta\tau_{\mathrm{a}}}{2})\leq |\vec{r}|^2\sin^2\!(\frac{\Theta}{2})
-x\sin^2(\Delta \tau_{\mathrm{a}}).
\end{cases}
\end{equation}
\end{theorem}
In this theorem, $\tau_{\mathrm{a}}$ is defined by
\begin{equation}
\tau_{\mathrm{a}}:=\frac{\Theta}{2\Delta\sqrt{1-\frac{4}{3}(|\vec{r}|^2-x-y)}}
\label{eq:max_1}
\end{equation}
for $x+y\geq |\vec{r}|^2-1/3$ and
\begin{equation}
\tau_{\mathrm{a}}:=\frac{\Theta}{2\Delta\sqrt{1\!-\!\left[\frac{1}{2}
+\sqrt{\frac{1}{3}(|\vec{r}|^2\!-\!x\!-\!y\!-\!\frac{1}{4})}\right]^{2}}}
\label{eq:max_2}
\end{equation}
for $x+y\leq |\vec{r}|^2-1/3$. The regime of the states given in the above
theorem is illustrated in the case of $|\vec{r}|^2=1$ and $\Theta=\pi/3$
[red (dark gray) area in Fig.~\ref{fig:3level_border}(a)]. For the states not in
this regime (in the gray area), there exist states for which the Bhatia-Davis formula
fails to be a valid lower bound. An interesting fact is that $\tau_{\mathrm{BD}}$
is a valid lower bound for most edges, for example, (1) $x=0$ and $y\neq 0$, i.e.,
nondiagonal states with $\rho_{12}=0$; (2) $x+y=|\vec{r}|^2$, which means $r_2=r_7=0$,
i.e., states with all diagonal entries $1/3$. The borderline between the red
(dark gray) and gray regimes reads
\begin{equation}
x\sin^2(\Delta \tau_{\mathrm{a}})+y\sin^2\left(\frac{\Delta\tau_{\mathrm{a}}}{2}\right)
=|\vec{r}|^2\sin^2\left(\frac{\Theta}{2}\right).
\label{eq:borderline}
\end{equation}
As shown in Fig.~\ref{fig:3level_border}(b), the area of violation regime (inside
the line) grows with the increase of $|\vec{r}|^2$ or the decrease of $\Theta$,
indicating that $\tau_{\mathrm{BD}}$ is a valid lower bound for most mixed states,
especially when $\Theta$ is large. As a matter of fact, apart from the case
$|\vec{r}|^2=1, \Theta=\pi/3$, this regime is very insignificant for other
examples given in Fig.~\ref{fig:S_3level}. Hence, in equally spaced three-level
systems, $\tau_{\mathrm{BD}}$ is indeed a valid lower bound for most states,
especially mixed states with large target angles.

\section{conclusion}

Inspired by the Bhatia-Davis theorem in mathematics and statistics, in this
paper we construct a formula, referred to as the Bhatia-Davis formula, for
the characterization of quantum speed limit in the Bloch representation. In
a general multilevel system, we first prove that the Bhatia-Davis formula is
an upper bound for the operational definition of quantum speed limit, and it
reduces to the operational definition when the average energy is half of the
summation between the maximum and minimum energies. The behaviors of both the
operational definition and Bhatia-Davis formula are discussed in the
generalized one-axis twisting model as an example. In the case of largest
target angle, the reachable state set $\mathcal{S}$ are first studied and the
Bhatia-Davis formula is then proved to be a valid lower bound for the evolution
time to reach the target in systems with symmetric energy structures.

With respect to few-level scenarios, the two-level systems are first studied,
and the Bhatia-Davis formula is proved to be a valid lower bound in this case,
and it reduces to the operational definition when attainable. Another alternative
state-dependent bound is also constructed using the Bhatia-Davis formula, which
is tighter than the bound given by the quantum Fisher information. In the case
of equally spaced three-level systems, the regime that the Bhatia-Davis formula
remains a valid lower bound is given. Even though it is not in general,
the violation becomes very insignificant for mixed states, especially when the
target angle is large. Therefore, it could be approximately treated as a valid
lower bound for most mixed states with large target angles in this type of systems.

\begin{acknowledgments}
The authors would like to thank M. Zhang and J. Qin for helpful discussions.
This work was supported by the National Natural Science Foundation of China
(Grants No.\,12175075, No.\,11805073, No.\,12088101, No.\,11935012, No.\,11875231,
and No.\,62003113), the NSAF (Grant No.\,U1930403), and the National Key Research
and Development Program of China (Grants No.\,2017YFA0304202 and No.\,2017YFA0205700).

J.L. and Z.M. contributed equally to this work.
\end{acknowledgments}

\appendix

\section{Calculation details for Theorem~\ref{theorem:upperbound} and its applications}
\label{sec:apx_tauBD_tau}

We first introduce the notation again for a better reading experience of the
appendix. $E_i$ is the $i$th energy of the Hamiltonian $H$ with the
corresponding eigenstate $|E_i\rangle$. Without loss of generality, here we
assume $E_0\leq E_1 \leq \cdots \leq E_{N-1}$ with $N$ the dimension of $H$.
In the energy basis $\{|E_i\rangle\}$, a density matrix can be expressed by
$\rho = \sum_{ij} \rho_{ij}|E_{i}\rangle\langle E_{j}|$, which immediately gives
$\langle H\rangle=\sum_i \rho_{ii}E_i$. Now define a function
\begin{equation}
f(x) := \sum_{i}\rho_{ii}(E_i-x)^2.
\end{equation}
It is easy to see that $x=\sum_{i}\rho_{ii}E_i=\langle H\rangle$ is the minimum value
of this function by calculating the first and second derivatives. Therefore, one can obtain
\begin{equation}
f\left(\frac{E_0+E_{N-1}}{2}\right)\geq f(\langle H\rangle).
\end{equation}
Next, by noticing that
\begin{eqnarray}
& & f\left(\frac{E_0+E_{N-1}}{2}\right)-f(\langle H\rangle) \nonumber \\
&=& \frac{1}{4}(E_{N-1}-E_0)^2\!-\!(E_{N-1}-\langle H\rangle)(\langle H\rangle-E_0),
\end{eqnarray}
one can obtain
\begin{equation}
\frac{1}{4}(E_{N-1}-E_0)^2\geq (E_{N-1}-\langle H\rangle)(\langle H\rangle-E_0),
\end{equation}
which leads to the result of our theorem below
\begin{equation}
\frac{\Theta}{2\sqrt{(E_{N-1}-\langle H\rangle)(\langle H\rangle-E_0)}}
\geq \frac{\Theta}{E_{N-1}-E_0},
\end{equation}
where $\Theta$ is the target angle and $\Theta/(E_{N-1}-E_0)$ is the OQSL for
time-independent Hamiltonians~\cite{Shao2020}. Theorem~\ref{theorem:upperbound}
is then proved.  \hfill $\blacksquare$

Now consider the generalized one-axis twisting Hamiltonian
\begin{equation}
H=\delta J_z+\chi J^2_z,
\end{equation}
where the angular momentum $J_z=\sum^n_{i=1}\sigma^{(i)}_{z}/2$ with
$\sigma^{(i)}_z$ the Pauli matrix along the $z$-axis for $i$th spin. The coherent
spin state can be expressed by
\begin{equation}
|\psi\rangle=e^{\zeta J_{+}-\zeta^{*} J_{-}}|J,J\rangle,
\end{equation}
where $J_{\pm}=J_x\pm i J_y$. Since $[J_{\pm},J_z]=\mp J_{\pm}$ and $[J_{+},J_{-}]=2J_z$,
one can obtain
\begin{eqnarray}
& & e^{-(\zeta J_{+}-\zeta^{*} J_{-})}J_ze^{\zeta J_{+}-\zeta^{*}J_{-}} \nonumber \\
&=& \cos(2|\zeta|)J_z+\frac{\sin(2|\zeta|)}{2|\zeta|}(\zeta J_{+}+\zeta^{*}J_{-}),
\end{eqnarray}
which immediately gives $\langle\psi| J_z|\psi\rangle= J\cos(2|\zeta|)$ and
\begin{equation}
\langle\psi| J^2_z|\psi\rangle=J^2\cos^2(2|\zeta|)+\frac{1}{2}J \sin^2(2|\zeta|).
\end{equation}
Hence, the expected value is
\begin{equation*}
\langle H\rangle=\delta J\cos(2|\zeta|)+\chi J^2\cos^2(2|\zeta|)
+\frac{\chi}{2}J \sin^2(2|\zeta|).
\end{equation*}
Due to the fact that $|\zeta|=\phi/2$ and $J=n/2$, the equation above
can be rewritten as
\begin{equation}
\langle H\rangle=\frac{1}{4}\left(2\delta n\cos\phi+\chi n^2\cos^2\phi
+\chi n \sin^2\phi\right).
\end{equation}

\section{Calculations and proofs in the case of largest target angle}
\label{sec:apx_multilevel}

For the time-independent Hamiltonians under unitary evolution, the set
$\mathcal{S}$ in the energy basis $\{|E_k\rangle\}$ can be written as~\cite{Shao2020}
\begin{eqnarray}
\mathcal{S} &=& \Bigg\{\vec{r}~\Big|1-\cos\Theta=\frac{1}{|\vec{r}|^2}\sum^{N-1}_{j=1}
\sum^{j-1}_{k=0}\left\{1-\cos\left[(E_j-E_k)t\right]\right\} \nonumber \\
& & \times \left(r^{2}_{j^{2}+2k-1}+r^{2}_{j^2+2k}\right), \exists t \Bigg\},
\label{eq:apx_S}
\end{eqnarray}
where $N$ is the dimension of Hamiltonian, $r_i$ is the $i$th entry of the Bloch
vector. Here the SU($N$) generators $\{\lambda_i\}$ are generated via the rules
\begin{equation}
\mathrm{(1)}~\lambda_{l^2-2}=\sqrt{\frac{2}{l(l-1)}}\mathrm{diag}(1,1,\dots,1-l,0,\dots)
\label{eq:apx_suN0}
\end{equation}
for $l=2,3,\dots,N$, namely, the first $l-1$ diagonal entries are 1, the $l$th
entry is $1-l$, and zero for others. (2) For $j\in[1,N-1]$ and $k\in(0,j-1]$,
the only nonzero entries in $\lambda_{j^2+2k-1}$ are the $kj$th and $jk$th ones,
and the corresponding values are 1. (3) For $j\in[1,N-1]$ and $k\in(0,j-1]$,
the only nonzero entries in $\lambda_{j^2+2k}$ are the $kj$th and $jk$th
ones, and the corresponding values are $-i$ and $i$, respectively. Specifically,
in the basis $\{E_k\}$ they can be expressed by
\begin{equation}
\lambda_{j^2+2k-1}=\left(\begin{array}{cccccc}
0 & \cdots & \cdots & \cdots &\cdots & 0\\
0 & 0 & \vdots & \vdots & \vdots & \vdots\\
\vdots & \cdots & \ddots & \cdots & 1 & \vdots\\
\vdots & 1 & \cdots &\ddots  & \cdots &\vdots\\
\vdots & \vdots & \vdots & \vdots & 0  & \vdots\\
0 & \cdots & \cdots & \cdots & \cdots & 0
\end{array}\right),
\label{eq:apx_suN1}
\end{equation}
and
\begin{equation}
\lambda_{j^2+2k}=\left(\begin{array}{cccccc}
0 & \cdots & \cdots & \cdots &\cdots & 0\\
0 & 0 & \vdots & \vdots & \vdots & \vdots\\
\vdots & \cdots & \ddots & \cdots & -i & \vdots\\
\vdots & i & \cdots &\ddots  & \cdots &\vdots\\
\vdots & \vdots & \vdots & \vdots & 0  & \vdots\\
0 & \cdots & \cdots & \cdots & \cdots & 0
\end{array}\right).
\label{eq:apx_suN2}
\end{equation}

In the case that the target $\Theta=\pi$, the constraint on $t$ in the equation
above reduces to
\begin{equation}
2=\sum^{N-1}_{j=1}\sum^{j-1}_{k=0}\left\{1-\cos\left[(E_j-E_k)t\right]\right\}
\frac{r^{2}_{j^{2}+2k-1}+r^{2}_{j^2+2k}}{|\vec{r}|^2}.
\label{eq:apx_targetPi}
\end{equation}
Since $1-\cos\left[(E_j-E_k)t\right]\leq 2$ and
\begin{equation}
\frac{1}{|\vec{r}|^2}\sum^{N-1}_{j=1}\sum^{j-1}_{k=0}r^{2}_{j^{2}+2k-1}
+r^{2}_{j^2+2k}\leq 1,
\label{eq:apx_targetPi_cond}
\end{equation}
the only solutions for Eq.~(\ref{eq:apx_targetPi}) are
\begin{equation}
(E_j-E_k)t=(2m+1)\pi,~~m=0,1,2,\cdots
\label{eq:apx_temp1}
\end{equation}
for all pairs of $j$ and $k$ that satisfy $r^2_{j^2+2k-1}+r^2_{j^2+2k}\neq 0$.
In the meantime, these solutions are valid only when the equality in
inequality~(\ref{eq:apx_targetPi_cond}) are attained, which also requires the
following additional condition:
\begin{equation}
r_{l^2-2}=0, \forall l=2,3,\cdots,N. \label{eq:apx_tauBD_condition}
\end{equation}
A useful fact is that in this case, regardless of the energy structures,
the state satisfying
\begin{equation}
r^2_{j^2+2k-1}+r^2_{j^2+2k}=|\vec{r}|^2
\end{equation}
for $j\in[1,N-1]$ and $k\in [0,j-1]$ can always reach the target $\Theta=\pi$ at
the time
\begin{equation}
t=\frac{\pi}{E_j-E_k}. \label{eq:apx_treach}
\end{equation}
Hence, $\mathcal{S}$ is not an empty set here, and in the meantime, the set
\begin{equation}
\mathcal{S}_0\!=\!\{\vec{r}\,|r^2_{j^2+2k-1}\!+\!r^2_{j^2+2k}\!=\!|\vec{r}|^2,
\forall j\!\in\![1,N\!-\!1],k\!\in\![0,j\!-\!1]\}
\end{equation}
is always a subset of $\mathcal{S}$. Theorem~\ref{theorem:multiB} is then proved.
\hfill $\blacksquare$

In fact, since $\lambda_{l^2-2}$ ($l=2,3,\dots,N$) is a diagonal SU($N$) generator,
Eq.~(\ref{eq:apx_tauBD_condition}) indicates that in the energy basis, the diagonal
entries of the density matrices which leads to valid solutions of $t$ must be
$1/N$. In the case of $N>2$ it cannot be pure states. Corollary~\ref{corollary:nopure}
is then prove.  \hfill $\blacksquare$

One should notice that whether Eq.~(\ref{eq:apx_temp1}) has more solutions apart
from Eq.~(\ref{eq:apx_treach}) depends on the energy structure. Recall that we
assume $E_0\leq E_1\leq\cdots\leq E_{N-1}$ and there exist at least two different
energies. Now denote $\mathcal{E}_{\mathrm{d}}$ as the set of all energy differences:
\begin{equation}
\mathcal{E}_{\mathrm{d}}=\{E_j-E_k|\,\forall j\in[1,N-1],k\in[0,j-1]\}.
\end{equation}
If the ratio between any two elements in $\mathcal{E}_{\mathrm{d}}$ cannot
be written in the form of $(2m_1+1)/(2m_2+1)$ with $m_1$, $m_2$ any two
non-negative integers, then only one pair of $(j,k)$ is allowed to satisfy
$r^2_{j^2+2k-1}+r^2_{j^2+2k}\neq 0$ to make sure Eq.~(\ref{eq:apx_temp1}) has
solutions, which means there are no other solutions except for Eq.~(\ref{eq:apx_treach}),
namely, $\mathcal{S}=\mathcal{S}_0$. One interesting specific example here is
that all the elements in $\mathcal{E}_{\mathrm{d}}$ are noncommensurable to
each other, which naturally fit the case that any two elements cannot be written
in the form of $(2m_1+1)/(2m_2+1)$.

Furthermore, due to the expressions of $\lambda_{j^2+2k-1}$ and $\lambda_{j^2+2k}$
given in Eqs.~(\ref{eq:apx_suN1}) and (\ref{eq:apx_suN2}), in the density matrix
representation, the states in $\mathcal{S}_0$ must take the form
\begin{equation}
\left(\begin{array}{cccccc}
\frac{1}{N} & \cdots & \cdots & \cdots & \cdots & 0 \\
\vdots & \frac{1}{N} & \vdots & \vdots & \vdots & \vdots\\
\vdots & \vdots & \ddots & \vdots & \rho_{kj} & \vdots\\
\vdots & \rho_{kj}^{*} & \vdots &\ddots  & \vdots & \vdots\\
\vdots & \vdots & \vdots & \vdots & \frac{1}{N}  & \vdots\\
0 & \cdots & \cdots & \cdots & \cdots & \frac{1}{N}
\end{array}\right),
\end{equation}
where all the diagonal entries are $1/N$, and all the nondiagonal entries are
zero except for the $kj$th and $jk$th ones. Corollaries~\ref{corollary:S0_1}
and~\ref{corollary:S0_2} are then proved. \hfill $\blacksquare$

In the following we continue to prove Theorem~\ref{theorem:tauBD_pi}. Since
the generators $\lambda_{l^2-2}$ for $l=2,3,\cdots,N$ are diagonal,
Eq.~(\ref{eq:apx_tauBD_condition}) indicates that
\begin{equation}
\vec{r}\cdot\mathrm{Tr}(H\vec{\lambda})
=\sum_{jk}r_j E_k\langle E_k|\lambda_j|E_k\rangle=0.
\end{equation}
This is due to the fact that for those nonzero entries of $\vec{r}$, the
corresponding SU($N$) generators have no nonzero diagonal entries. Therefore, the
average energy reads
\begin{equation}
\langle H\rangle=\frac{1}{N}\mathrm{Tr}(H)=\frac{1}{N}\sum^{N-1}_{k=0}E_k.
\end{equation}
In the case that the energies are symmetric about $(E_0+E_{N-1})/2$,
\begin{equation}
E_{N-1}+E_0 = E_{N-1-k}+E_k \label{eq:apx_symmetry}
\end{equation}
for any subscript $k$ satisfying $E_k\leq (E_0+E_{N-1})/2$, the average
energy further reduces to
\begin{equation}
\langle H\rangle=\frac{1}{2}(E_0+E_{N-1}),
\end{equation}
and the Bhatia-Davis formula reduces to
\begin{equation}
\tau_{\mathrm{BD}}=\frac{\pi}{E_{N-1}-E_0},
\end{equation}
which is nothing but the OQSL~\cite{Shao2020}. Hence, $\tau_{\mathrm{BD}}$ is
a valid lower bound in this case. For the states not in $\mathcal{S}$, the target
cannot be fulfilled, meaning that the time is infinite, and $\tau_{\mathrm{BD}}$
is also a valid formal lower bound. Thus, for a symmetric spaced Hamiltonian,
$\tau_{\mathrm{BD}}$ is a valid state-dependent lower bound for $\Theta=\pi$.
Theorem~\ref{theorem:tauBD_pi} is then proved. \hfill $\blacksquare$

Moreover, Eq.~(\ref{eq:apx_symmetry}) indicates that
\begin{equation}
\frac{E_{N-1-k}-E_{N-1-m}}{E_m-E_k}=1
\end{equation}
for $k<j\leq\lfloor\frac{N-1}{2}\rfloor$ with $\lfloor\cdot\rfloor$ the floor
function, which means Eq.~(\ref{eq:apx_temp1}) for the pairs of subscripts $j,k$
and $N-1-k,N-1-j$ can always hold simultaneously. Hence, the states satisfying
\begin{eqnarray}
|\vec{r}|^2 & = & r^2_{j^2+2k-1}+r^2_{j^2+2k}+r^2_{(N-1-k)^2+2(N-1-j)-1} \nonumber \\
& & +r^2_{(N-1-k)^2+2(N-1-j)}
\end{eqnarray}
can always fulfill the target $\Theta=\pi$ in this case. In the density matrix
representation, due to Eqs.~(\ref{eq:apx_suN1}) and (\ref{eq:apx_suN2}), these
state are of the form
\begin{equation*}
\left(\begin{array}{ccccccc}
\frac{1}{N} & 0 & \cdots & \cdots &\cdots & \cdots & 0\\
0 & \frac{1}{N} & \vdots & \vdots &\vdots & \vdots & \vdots\\
\vdots & \vdots & \ddots & \rho_{kj} & \vdots & \vdots & \vdots\\
\vdots & \vdots & \rho^*_{kj} & \ddots & \rho_{N-1-j,N-1-k} & \vdots & \vdots\\
\vdots & \vdots & \vdots & \rho^*_{N-1-j,N-1-k} & \ddots & \vdots & \vdots\\
\vdots & \vdots & \vdots & \vdots & \vdots  & \ddots & \vdots\\
0 & \cdots & \cdots & \cdots & \cdots & \cdots & \frac{1}{N}
\end{array}\right).
\end{equation*}

\section{Calculation details in two-level systems}
\label{sec:apx_twolevel}

Now we prove that the Bhatia-Davis formula is a valid lower bound in two-level
systems under unitary evolution. For a two-level system, the Bloch vector of a
state can be expressed by
\begin{equation}
\vec{r}=\eta(\sin\alpha\cos\varphi,\sin\alpha\sin\varphi,\cos\alpha)^{\mathrm{T}},
\end{equation}
where $\eta\in(0,1]$, $\alpha\in [0,\pi]$, $\varphi\in[0,2\pi]$ and the superscript
T represents the transposition. In this case, the set $\mathcal{S}$ reads~\cite{Shao2020}
\begin{equation}
\mathcal{S}=\left\{\vec{r}~\Big|\eta\in(0,1],\alpha\in\left[\frac{\Theta}{2},
\pi-\frac{\Theta}{2}\right]\!, \varphi\in[0,2\pi]\right\}\!.
\end{equation}
From the analysis in Appendix D of Ref.~\cite{Shao2020}, one can see that
for $\alpha\leq \pi/2$, the evolution time for the states in $\mathcal{S}$ to
reach the target angle $\Theta$ can be expressed by
\begin{equation}
t=\frac{2}{E_1-E_0}\arcsin\left(\frac{\sin(\frac{\Theta}{2})}{\sin\alpha}\right),
\end{equation}
for $\alpha>\Theta/2$, and $t=\pi/(E_1-E_0)$ for $\alpha=\Theta/2$. Furthermore,
$\tau_{\mathrm{BD}}$ in this case can be calculated as
\begin{equation}
\tau_{\mathrm{BD}}=\frac{\Theta}{(E_{1} - E_{0})\sqrt{1-\eta^{2}\cos^{2}\alpha}},
\end{equation}
from which one can see that $\tau_{\mathrm{BD}}$ is a monotonically increasing
function with respect to $|\vec{r}|$:
\begin{equation}
\tau_{\mathrm{BD}} \leq \frac{\Theta}{(E_{1}-E_{0})\sin{\alpha}}.
\label{eq:inter0}
\end{equation}
From the trigonometric inequality $\sin x\leq x$, it is easy to see
$x\leq \arcsin x$. Utilizing this inequality, we have
\begin{equation}
\frac{\sin\left(\frac{\Theta}{2}\right)}{\sin\alpha}\leq
\arcsin\left(\frac{\sin(\frac{\Theta}{2})}{\sin\alpha}\right),
\end{equation}
which immediately gives
\begin{equation}
t\geq \frac{\Theta}{(E_{1}-E_{0})\sin{\alpha}} \geq \tau_{\mathrm{BD}}.
\end{equation}
Furthermore, one should notice that $\tau_{\mathrm{BD}}$ can be attained when
$\alpha=\pi/2$. The case of $\alpha>\pi/2$ can be analyzed similarly due to the
symmetry of the dynamics. For the states out of $\mathcal{S}$, the target
angle cannot be reached, indicating that any finite value could be a mathematically
valid lower bound for it. Therefore, $\tau_{\mathrm{BD}}$ is also a valid bound
in this regime. Theorem~\ref{theorem:tauBD_2level} is then proved. \hfill $\blacksquare$

Recall that $\tau_{\mathrm{F}}=2\mathcal{A}/\sqrt{F}$ and $\tau_{\mathrm{C}}
=\max\left\{\frac{\mathcal{A}}{\Delta H}, \frac{2\mathcal{A}^2}
{\pi\langle H\rangle}\right\}$, now we prove that $\tau_{\mathrm{F}}\geq
\tau_{\mathrm{C}}$ for two-level systems. First, through some straightforward
calculations (more mathematical technologies for the calculation of quantum Fisher
information and quantum Fisher information matrix can be found in Refs.~\cite{Toth2014,
Liu2020}), one can obtain that
\begin{equation}
\tau_{\mathrm{F}}=\frac{2\mathcal{A}}{(E_1-E_0)\sqrt{|\vec{r}|^2-r^2_z}}
=\frac{2\mathcal{A}}{(E_1-E_0)\eta\sin\alpha}.
\end{equation}
Also,
\begin{equation}
\frac{\mathcal{A}}{\Delta H}=\frac{2\mathcal{A}}{(E_1-E_0)\sqrt{1-\eta^2\cos^2\alpha}},
\end{equation}
and
\begin{eqnarray}
\frac{2\mathcal{A}^2}{\pi \langle H\rangle}&=&\frac{2\mathcal{A}^2/\pi}
{\frac{1}{2}(E_1+E_0)+\frac{1}{2}\eta\cos\alpha(E_1-E_0)} \nonumber \\
&=&\frac{4\mathcal{A}^2/\pi}{(E_1-E_0)\left(\frac{E_1+E_0}{E_1-E_0}
+\eta\cos\alpha\right)}.
\end{eqnarray}
It is easy to see that $\eta\sin\alpha<\sqrt{1-\eta^2\cos^2\alpha}$, hence,
$\tau_{\mathrm{F}}\geq \mathcal{A}/\Delta H$. As a matter of fact, in the case
of two-level systems, $\mathcal{A}=\arccos\sqrt{1-\frac{1}{2}|\vec{r}|^2
(1-\cos\Theta)}\leq\Theta/2$, which means $\tau_{\mathrm{BD}}$ is also larger
than $\mathcal{A}/\Delta H$. Next, due to the fact $2\mathcal{A}<\pi$, one can
have
\begin{equation}
\frac{2\mathcal{A}^2}{\pi \langle H\rangle}\leq\frac{2\mathcal{A}}
{(E_1-E_0)\left(\frac{E_1+E_0}{E_1-E_0}+\eta\cos\alpha\right)}.
\label{eq:apxqubit_Ml}
\end{equation}
Hence, when
\begin{equation}
\frac{E_1+E_0}{E_1-E_0}\geq \sqrt{2}\eta
\end{equation}
is satisfied, the right-hand term of inequality~(\ref{eq:apxqubit_Ml}) is less
than $\tau_{\mathrm{F}}$, indicating that $\tau_{\mathrm{F}}\geq \tau_{\mathrm{C}}$
in this case.

\section{Calculation details in three-level systems}
\label{sec:apx_3level}

\subsection{Conditions for Bloch vectors}
\label{sec:apx_3level1}

In the case of three-level systems, the density matrix $\rho$ can be expressed by
\begin{equation}
\rho=\frac{1}{3}\left(\openone+\sqrt{3}\vec{r}\cdot\vec{\lambda}
\right),
\label{eq:apx_rho}
\end{equation}
where $\Vec{r}=(r_0,r_1,r_2,r_3,r_4,r_5,r_6,r_7)^{\mathrm{T}}$ is the Bloch
vector, and the specific form of SU(3) generators in the energy basis
$\{|E_0\rangle,|E_1\rangle,|E_2\rangle \}$ ($E_0\leq E_1\leq E_2$) given in
Eqs.~(\ref{eq:apx_suN0}), (\ref{eq:apx_suN1}), and (\ref{eq:apx_suN2}) are nothing
but the following Gell-Mann matrices:
\begin{align}
\lambda_0 &=
\left(\begin{array}{ccc}
0 & 1 & 0 \\
1 & 0 & 0 \\
0 & 0 & 0
\end{array}\right), ~~
\lambda_1 =
\left(\begin{array}{ccc}
0 & -i & 0 \\
i & 0 & 0 \\
0 & 0 & 0
\end{array}\right), \\
\lambda_2 &=
\left(\begin{array}{ccc}
1 & 0 & 0 \\
0 & -1 & 0 \\
0 & 0 & 0
\end{array}\right), ~~
\lambda_3 =
\left(\begin{array}{ccc}
0 & 0 & 1 \\
0 & 0 & 0 \\
1 & 0 & 0
\end{array}\right), \\
\lambda_4 &=
\left(\begin{array}{ccc}
0 & 0 & -i \\
0 & 0 & 0 \\
i & 0 & 0
\end{array}\right), ~~
\lambda_5 =
\left(\begin{array}{ccc}
0 & 0 & 0 \\
0 & 0 & 1 \\
0 & 1 & 0
\end{array}\right), \\
\lambda_6 &=
\left(\begin{array}{ccc}
0 & 0 & 0 \\
0 & 0 & -i \\
0 & i & 0
\end{array}\right), ~~
\lambda_7 =
\frac{1}{\sqrt{3}}\left(\begin{array}{ccc}
1 & 0 & 0 \\
0 & 1 & 0 \\
0 & 0 & -2
\end{array}\right).
\end{align}
Substituting $\vec{r}$ and Gell-Mann matrices into Eq.~(\ref{eq:apx_rho}),
the density matrix can be written as
\begin{equation*}
\left(\begin{array}{ccc}
\frac{1}{3}+\frac{1}{3}r_7+\frac{1}{\sqrt{3}}r_2 &
\frac{1}{\sqrt{3}}r_0-\frac{i}{\sqrt{3}}r_1 &
\frac{1}{\sqrt{3}}r_3-\frac{i}{\sqrt{3}}r_4 \\
\frac{1}{\sqrt{3}}r_0+\frac{i}{\sqrt{3}}r_1 &
\frac{1}{3}+\frac{1}{3}r_7-\frac{1}{\sqrt{3}}r_2 &
\frac{1}{\sqrt{3}}r_5-\frac{i}{\sqrt{3}}r_6  \\
\frac{1}{\sqrt{3}}r_3+\frac{i}{\sqrt{3}}r_4 &
\frac{1}{\sqrt{3}}r_5+\frac{i}{\sqrt{3}}r_6 &
\frac{1}{3}-\frac{2}{3}r_7
\end{array}\right)\!\!.
\end{equation*}

Since $\rho$ is positive semidefinite, due to the Schur complement theorem
one could have (1) the matrix
\begin{equation}
A = \left(\begin{array}{cc}
\frac{1}{3}+\frac{1}{3}r_7+\frac{1}{\sqrt{3}}r_2 & \frac{1}{\sqrt{3}}r_0-\frac{i}{\sqrt{3}}r_1 \\
\frac{1}{\sqrt{3}}r_0+\frac{i}{\sqrt{3}}r_1 & \frac{1}{3}+\frac{1}{3}r_7-\frac{1}{\sqrt{3}}r_2
\end{array}\right)
\end{equation}
is positive semidefinite and (2) the Schur complement
\begin{equation}
\frac{1}{3}-\frac{2}{3}r_7-\vec{v}^{\,\dagger}A^{-1}\vec{v}\geq 0
\label{eq:apx_sc}
\end{equation}
with $\vec{v}=\frac{1}{\sqrt{3}}(r_3-i r_4, r_5-i r_6)^{\mathrm{T}}$.
Since the eigenvalues of A are $\frac{1}{3}\big[1+r_{7}\pm\sqrt{3(r_{0}^{2}+r_{1}^{2}
+r_{2}^{2})}\,\big]$, the positive semidefinite property indicates
\begin{equation}
\frac{1}{\sqrt{3}}(1+r_{7})\geq \sqrt{r_{0}^{2}+r_{1}^{2}+r_{2}^{2}}.
\label{eq:apx_Schur}
\end{equation}
Also, from the expression of the density matrix, to guarantee all diagonal
entries non-negative, i.e., $\frac{1}{3}+\frac{1}{3}r_7+\frac{1}{\sqrt{3}}
r_2\leq 1$ and $\frac{1}{3}-\frac{2}{3}r_7\geq 0$, $r_2$ and $r_7$ need to satisfy
\begin{equation}
r_7\leq \frac{1}{2},~~|r_2|\leq\frac{\sqrt{3}}{2}.
\end{equation}

\subsection{Calculation of $\mathcal{S}$ in equally spaced three-level
systems}

In the case of equally spaced Hamiltonian, the constraint in Eq.~(\ref{eq:apx_S})
reduces to
\begin{equation}
2x\cos^2(\Delta t)+y\cos(\Delta t)-2x-y+|\vec{r}|^2(1-\cos\Theta)\!=\!0,
\label{eq:S_constrain}
\end{equation}
where
\begin{align}
x &= r^2_3+r^2_4, \\
y &= r^2_0+r^2_1+r^2_5+r^2_6.
\end{align}
It is easy to see that $x$, $y$ are both non-negative and satisfy
\begin{equation}
x+y\leq |\vec{r}|^2.
\label{eq:apx_normalization}
\end{equation}
The search of $\mathcal{S}$ is equivalent to the search of regimes of $x$ and
$y$ [together with Eqs.~(\ref{eq:apx_sc}) and (\ref{eq:apx_Schur})] that allow
reasonable solutions of $t$. It is not difficult to see that there is no solution
for $t$ when $x=y=0$. Hence, the discussion can be divided into two parts, (1)
$x=0$ ($y\neq 0$) and (2) $x\neq 0$. Now we discuss them individually.

In case (1) $x=0$ ($y\neq 0$), Eq.~(\ref{eq:S_constrain}) reduces to
\begin{equation}
\cos(\Delta t)=1-\frac{2}{y}|\vec{r}|^2(1-\cos\Theta).
\label{eq:apx_x0}
\end{equation}
The right-hand term is naturally not larger than 1, and requiring it to be not less
than $-1$ immediately leads to $y\geq\frac{1}{2}|\vec{r}|^2(1-\cos\Theta)
=|\vec{r}|^2\sin^2(\frac{\Theta}{2})$. Therefore, a feasible regime for legitimate
solutions of $t$ is
\begin{equation}
x=0,~y\in\left[|\vec{r}|^2\sin^2\left(\frac{\Theta}{2}\right),~|\vec{r}|^2\right].
\end{equation}

In case (2) $x\neq 0$, the formal solution for Eq.~(\ref{eq:S_constrain}) is
\begin{equation}
\cos{(\Delta t)}\!=\!\frac{-y\!\pm\!\sqrt{(4x\!+\!y)^{2}\!-\!16x|\vec{r}|^2
\sin^2(\frac{\Theta}{2})}}{4x}\!:=\!f_{\pm}.
\label{eq:apx_cosdt}
\end{equation}
To make sure there exist legitimate solutions, the first requirement is
$(4x+y)^{2}\geq 8x|\vec{r}|^2(1-\cos\Theta)$:
\begin{equation}
y \geq 4|\vec{r}|\sin\left(\frac{\Theta}{2}\right)\sqrt{x}-4x.
\label{eq:apx_cond1}
\end{equation}
In the case that $y=0$, it reduces to $x\geq |\vec{r}|^2\sin^2(\frac{\Theta}{2})$.
Hence, on the $x$ axis, the feasible regime for legitimate solutions of $t$ is
\begin{equation}
x\in\left[|\vec{r}|^2\sin^2\left(\frac{\Theta}{2}\right),~|\vec{r}|^2\right],
~y=0.
\end{equation}

The second requirement is that at least one of the conditions $f_{+}\in [-1,1]$,
$f_{-}\in [-1,1]$ can be satisfied. Due to the fact that $f_{\pm}\leq 1$ is always
satisfied, we need to consider only the case that at least one of $f_{+}\geq -1$
and $f_{-}\geq -1$ is satisfied. Since $f_{\pm}\geq -1$ can be rewritten into
\begin{equation}
y\leq 4x\pm\sqrt{(4x+y)^2-16x|\vec{r}|^2\sin^2\left(\frac{\Theta}{2}\right)}.
\end{equation}
For the sake of the largest regime, we need to take only the positive sign one:
\begin{equation}
y\leq 4x+\sqrt{(4x+y)^2-16x|\vec{r}|^2\sin^2\left(\frac{\Theta}{2}\right)}.
\end{equation}
When $y\leq 4x$, this inequality is naturally satisfied. When $y\geq 4x$, it
reduces to
\begin{equation}
y\geq |\vec{r}|^2\sin^2\left(\frac{\Theta}{2}\right).
\end{equation}
In a word, the second requirement can be rewritten into $y\leq 4x$ or
\begin{equation}
\begin{cases}
y\geq 4x, \\
y\geq |\vec{r}|^2\sin^2\left(\frac{\Theta}{2}\right).
\end{cases}
\end{equation}

Combing both requirements, the conditions for $x$ and $y$ that guarantee legitimate
solutions of $t$ are
\begin{equation}
\begin{cases}
y \geq 4|\vec{r}|\sin\left(\frac{\Theta}{2}\right)\sqrt{x}-4x, \\
y\leq 4x, \\
y\leq |\vec{r}|^2-x.
\end{cases} \label{eq:apx_regime1}
\end{equation}
and
\begin{equation}
\begin{cases}
y \geq 4|\vec{r}|\sin\left(\frac{\Theta}{2}\right)\sqrt{x}-4x, \\
y\geq 4x, \\
y\geq |\vec{r}|^2\sin^2\left(\frac{\Theta}{2}\right), \\
y\leq |\vec{r}|^2-x.
\end{cases} \label{eq:apx_regime2}
\end{equation}

\subsection{Proof of Theorem~\ref{lamma_threelevel}}

In the case of three-level systems, the Bhatia-Davis formula reads
\begin{equation}
\tau_{\mathrm{BD}}=\frac{\Theta}{2\sqrt{(E_2-\langle H\rangle)
(\langle H\rangle-E_{0})}}. 
\end{equation}
In the energy basis $\{|E_0\rangle,|E_1\rangle,|E_2\rangle\}$,
the expected value of $H$ is
\begin{eqnarray}
\langle H \rangle &=& \frac{1}{3}(E_{0}+E_{1}+E_{2})
+\frac{1}{\sqrt{3}}r_{2}(E_{0}-E_{1}) \nonumber \\
& & +\frac{1}{3}r_{7}(E_{0}+E_{1}-2E_{2}).
\label{eq:apx_H}
\end{eqnarray}
In the case that the energy levels are equally spaced, i.e.,
$E_{1}= E_{0}+\Delta$ and $E_{2}=E_{0}+2\Delta$ with $\Delta$ a constant gap,
$\langle H\rangle$ reduces to
\begin{equation}
\langle H\rangle= E_{0}+\Delta\left(1-\frac{1}{\sqrt{3}}r_{2}-r_{7}\right),
\end{equation}
which directly gives that
\begin{equation}
\tau_{\mathrm{BD}}=\frac{\Theta}
{2\Delta\sqrt{1-\left(\frac{1}{\sqrt{3}}r_{2}+r_{7}\right)^{2}}},
\end{equation}
and
\begin{equation}
\cos(\Delta\tau_{\mathrm{BD}})=\cos\left(\frac{\Theta}
{2\sqrt{1-\left(\frac{1}{\sqrt{3}}r_{2}+r_{7}\right)^{2}}}\right).
\label{eq:apx_costauBD}
\end{equation}

To prove Theorem~\ref{lamma_threelevel}, we need to calculate the maximum
value of $\frac{1}{\sqrt{3}}r_2+r_7$ for states in set $\mathcal{S}$. The
mathematical statement of this problem is
\begin{align}
&\max~~\frac{1}{\sqrt{3}}r_2+r_7 \nonumber \\
&\mathrm{subject~to~~(\ref{eq:apx_regime1})~and~(\ref{eq:apx_regime2})},
\end{align}
for a fixed $|\vec{r}|^2$ and $\Theta$.

\begin{figure}[tp]
\includegraphics[width=8.5cm]{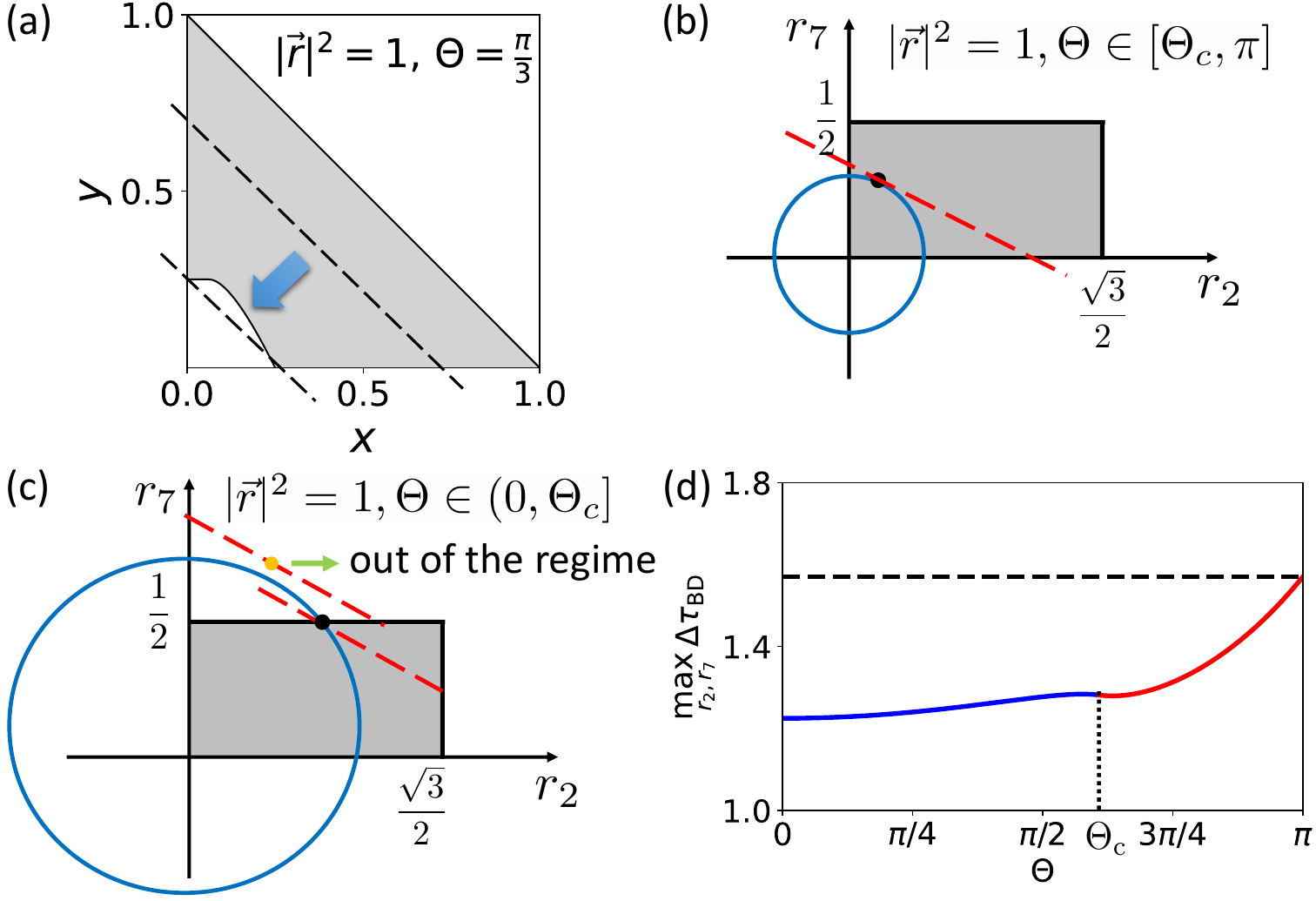}
\caption{Illustration of the proof of Theorem~\ref{lamma_threelevel}, including
(a) minimization of $x+y$, (b) maximization of $\frac{1}{\sqrt{3}}r_2+r_7$ for
$|\vec{r}|^2=1$ and $\Theta\in [\Theta_{\mathrm{c}},\pi]$, (c) maximization of
$\frac{1}{\sqrt{3}}r_2+r_7$ for $|\vec{r}|^2=1$ and $\Theta\in (0,\Theta_{\mathrm{c}}]$,
and (d) $\max_{r_2,r_7}\Delta\tau_{\mathrm{BD}}$ as a function of target $\Theta$.
Here $\Theta_{\mathrm{c}}=2\arccos(1/\sqrt{3})$.}
\label{fig:apx_proof0}
\end{figure}

Conditions~(\ref{eq:apx_regime1})~and~(\ref{eq:apx_regime2}) are not direct
constraints on $r_2$ and $r_7$, but on $x$ and $y$. Since $r^2_2+r^2_7=|\vec{r}|^2-x-y$,
these constraints can affect only the value of $r^2_2+r^2_7$. Hence, the first
optimization in this case is to minimize $x+y$, which corresponds to the maximum
$r^2_2+r^2_7$. In the regime given by inequalities (\ref{eq:apx_regime1}) and
(\ref{eq:apx_regime2}), as illustrated in Fig.~\ref{fig:apx_proof0}(a) (we
take these specific values of $|\vec{r}|^2$ and $\Theta$ only for demonstration;
the calculation is valid for any value), different values of $x+y$ mean different
position of the dashed black line in the plot. It is obvious that the minimum
value is attained when the line is closest to the original point, which gives
$\min\,x+y=|\vec{r}|^2\sin^2(\frac{\Theta}{2})$:
\begin{equation}
\max_{x,y}\,r^2_2+r^2_7=|\vec{r}|^2\cos^2\left(\frac{\Theta}{2}\right).
\end{equation}
The expression above can be further optimized with respect to $|\vec{r}|^2$, i.e.,
$\max_{|\vec{r}|^2}\,r^2_2+r^2_7=\cos^2(\frac{\Theta}{2})$. Next, we need to
optimize the value $\frac{1}{\sqrt{3}}r_2+r_7$ with the constraint $r^2_2+r^2_7=
\cos^2(\frac{\Theta}{2})$. Due to the semidefinite property of density matrix
discussed in Appendix~\ref{sec:apx_3level1}, there we have $r_{7}\leq 1/2$
and $r_2\leq\sqrt{3}/2$, as the gray areas shown in Figs.~\ref{fig:apx_proof0}(b)
and \ref{fig:apx_proof0}(c). Also, the constraint equation $r^2_2+r^2_7=\cos^2(\frac{\Theta}{2})$
is a circle (blue circles in the plots). The dashed red line represents $\frac{1}{\sqrt{3}}
r_2+r_7=c$ with $c$ a constant. Regardless of the constraints $r_{7}\leq 1/2$
and $r_2\leq \sqrt{3}/2$, the maximum value of $\frac{1}{\sqrt{3}}r_2+r_7$
is always attained (denoted by the black dot) with the dashed red line being the
tangent line of the circle. The values of $r_2$ and $r_7$ for this crossover
point are
\begin{equation}
r_2=\frac{1}{2}\cos\left(\frac{\Theta}{2}\right),~
r_7=\frac{\sqrt{3}}{2}\cos\left(\frac{\Theta}{2}\right).
\end{equation}

Now let us take into account the constraint on $r_2$ and $r_7$. For large
values of $\Theta$, the crossover point stays in the gray area, as shown in
Fig.~\ref{fig:apx_proof0}(b), then the maximum value of $\frac{1}{\sqrt{3}}r_2+r_7$
is attained by this point. For small values of $\Theta$, it is possible that the
value of $r_7$ for this point [orange point in Fig.~\ref{fig:apx_proof0}(c)]
is out of the gray area. In this case the maximum value is attained by the
point on the circle with $r_7=1/2$. The critical target $\Theta_{\mathrm{c}}$
satisfies $\frac{\sqrt{3}}{2} \cos\left(\frac{\Theta_{\mathrm{c}}}{2}\right)=\frac{1}{2}$,
which gives $\Theta_{\mathrm{c}}=2\arccos(\frac{1}{\sqrt{3}})$. Thus, in the case of
$\Theta\in[\Theta_{\mathrm{c}},\pi]$, $\max\,\Delta\tau_{\mathrm{BD}}$ reads
\begin{equation}
\max_{r_2,r_7}\,\Delta\tau_{\mathrm{BD}}=\frac{\sqrt{3}\Theta}{2\sqrt{1-2\cos\Theta}},
\label{eq:apx_maxutauBD_a}
\end{equation}
and in the case $\Theta\in(0, \Theta_{\mathrm{c}}]$, it is
\begin{equation}
\max_{r_2,r_7}\,\Delta\tau_{\mathrm{BD}}=\frac{\Theta}{2\sqrt{1-\frac{1}{4}
\left[\sqrt{\frac{1}{3}(1+2\cos\Theta)}+1\right]^2}}.
\label{eq:apx_maxutauBD_b}
\end{equation}
Both Eqs.~(\ref{eq:apx_maxutauBD_a}) and (\ref{eq:apx_maxutauBD_b}) are plotted
in Fig.~\ref{fig:apx_proof0}(d) as a function of $\Theta$. It can be seen
that the maximum value with respect to $\Theta$ is attained at $\Theta=\pi$,
and the corresponding value of $\Delta\tau_{\mathrm{BD}}$ is $\pi/2$. Hence,
one can obtain that
\begin{equation}
\Delta\tau_{\mathrm{BD}} \leq \frac{\pi}{2}
\end{equation}
for any legitimate values of $|\vec{r}|^2$ and $\Theta$. Theorem~\ref{lamma_threelevel}
is then proved. \hfill $\blacksquare$

\subsection{Analysis and proof of Theorem~\ref{theorem:3level}}

\begin{figure}[tp]
\includegraphics[width=8.5cm]{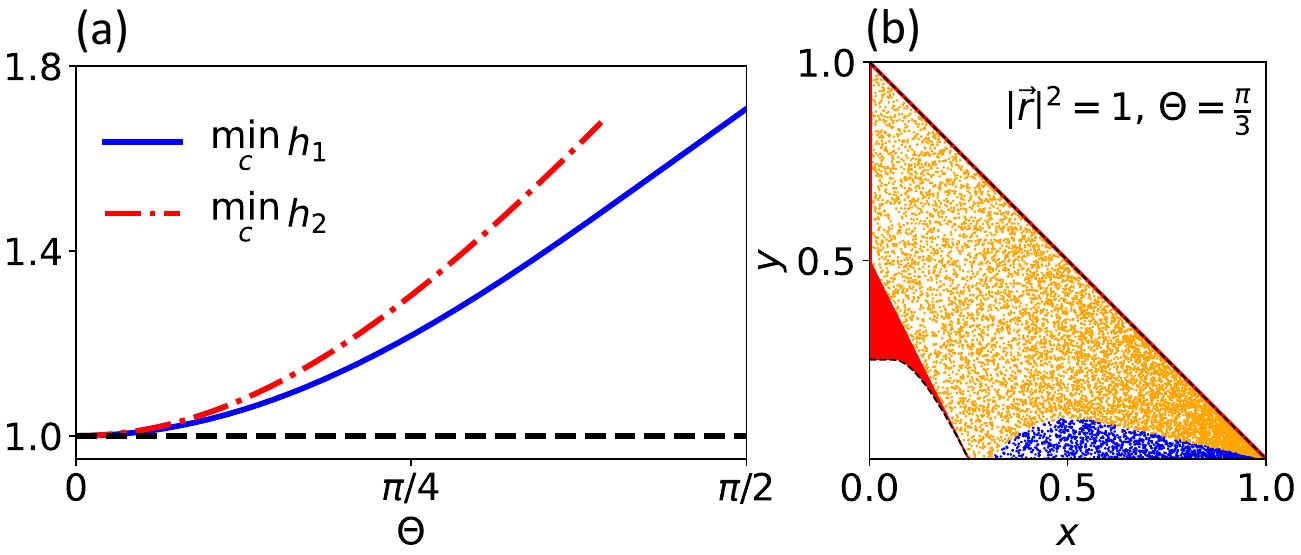}
\caption{(a) Minimum values of $h_1$ and $h_2$ with respect to $c$
($\min_{c}h_1$ and $\min_{c}h_2$) as a function of $\Theta$. (b) Numerical test
of the violation of $t\geq\tau_{\mathrm{BD}}$ in the case $|\vec{r}|^2=1$ and
$\Theta=\pi/3$. The blue and orange dots represent the states that
$t\geq\tau_{\mathrm{BD}}$ is violated or not. The red (dark gray) areas in
online (print) version are the regimes that $t\geq\tau_{\mathrm{BD}}$ is
guaranteed to hold.
\label{fig:apx_3levelproof}}
\end{figure}

With Theorem~\ref{lamma_threelevel}, we need to consider only the solutions of
Eqs.~(\ref{eq:apx_x0}) and (\ref{eq:apx_cosdt}) within the regime $(0,\frac{\pi}{2}]$
as the solutions not in this regime are obviously larger than $\tau_{\mathrm{BD}}$.
Now we compare the values of $\cos(\Delta t)$ in Eqs.~(\ref{eq:apx_x0}) and
(\ref{eq:apx_cosdt}) with $\cos(\Delta\tau_{\mathrm{BD}})$. We first consider
Eq.~(\ref{eq:apx_x0}), i.e., $x=0$. In this case, when $\Theta\geq\frac{\pi}{2}$,
there is
\begin{equation}
\frac{1}{y}|\vec{r}|^2(1-\cos\Theta)=
\frac{|\vec{r}|^2(1-\cos\Theta)}{|\vec{r}|^2-r^2_2-r^2_7}\geq 1,
\end{equation}
indicating that $\cos(\Delta t)\leq 0$. Therefore, $t\geq \tau_{\mathrm{BD}}$
in this case. When $\Theta <\pi/2$, if $r^2_2+r^2_7\in[|\vec{r}|^2\cos\Theta,
\frac{1}{2}|\vec{r}|^2(1+\cos\Theta)]$, $\cos(\Delta t)$ is also negative and
then $t\geq\tau_{\mathrm{BD}}$. Therefore, the only problem left is that if
this inequality is still valid for $r^2_2+r^2_7\in[0,|\vec{r}|^2\cos\Theta]$.
That is equivalent to proving the expression
\begin{equation}
\cos\left(\frac{\Theta}{2\sqrt{1-\left(\frac{1}{\sqrt{3}}r_{2}+r_{7}\right)^{2}}}\right)
+\frac{|\vec{r}|^2(1-\cos\Theta)}{|\vec{r}|^2-r^2_2-r^2_7}
\label{eq:apx_3level1}
\end{equation}
is larger than 1. In this expression, the minimum value with respect to
$|\vec{r}|^2$ is attained at $|\vec{r}|^2=1$. With this condition, we further use
a two-step method to locate the minimum value of the expression above. The first
step is to optimize $\frac{1}{\sqrt{3}}r_2+r_7$ with a fixed value of $r^2_2+r^2_7$,
i.e., $r^2_2+r^2_7=c$ with $c$ a constant. Then the second step is to optimize $c$.
In the first step, when $\cos\Theta\leq\frac{1}{3}$, the maximum value of
$\frac{1}{\sqrt{3}}r_2+r_7$ is $\frac{2\sqrt{c}}{\sqrt{3}}$, which is attained
by the tangent line on the circle with $r_2=\frac{\sqrt{c}}{2}$ and $r_7=\frac{\sqrt{3c}}{2}$,
as discussed in the proof of Theorem~\ref{lamma_threelevel}. In this case,
the minimum value of Eq.~(\ref{eq:apx_3level1}) reduces to
\begin{equation}
\cos\left(\frac{\Theta}{2\sqrt{1-\frac{4}{3}c}}\right)
+\frac{1-\cos\Theta}{1-c}:=h_1.
\label{eq:apx_3level_temp1}
\end{equation}
When $\cos\Theta>\frac{1}{3}$, the expression still keeps to be the one above
for the case $c\in[0,\frac{1}{3}]$. Whereas for $c\in[\frac{1}{3},\cos\Theta]$,
the maximum value of $\frac{1}{\sqrt{3}}r_2+r_7$ is attained by $r_{2}=\sqrt{c-\frac{1}{4}}$
and $r_7=\frac{1}{4}$, and the minimum value of Eq.~(\ref{eq:apx_3level1})
reduces to
\begin{equation}
\cos\!\left(\frac{\Theta}{2\sqrt{1\!-\!\left[\frac{1}{2}
\!+\!\sqrt{\frac{1}{3}(c\!-\!\frac{1}{4})}\right]^{2}}}\right)
\!+\!\frac{1\!-\!\cos\Theta}{1\!-\!c}:=h_2.
\label{eq:apx_3level_temp2}
\end{equation}
The minimum value of $h_1$ and $h_2$ with respect to $c$, i.e., $\min_{c}h_1$
(solid blue line) and $\min_{c}h_2$ (dash-dotted red line), are given in
Fig.~\ref{fig:apx_3levelproof}(a) as a function of $\Theta$. It can be seen
that in both cases the minimum values for any $\Theta$ is not smaller than one
(dashed black line). Therefore, Eq.~(\ref{eq:apx_3level1}) is indeed always
larger than 1 and $t\geq\tau_{\mathrm{BD}}$ holds here.

In the case that $x\neq 0$, we need to compare Eq.~(\ref{eq:apx_cosdt}) with
$\cos(\Delta \tau_{\mathrm{BD}})$.  It is obvious that the solution $f_{-}$ is
negative and the corresponding $\Delta t$ is larger than $\pi/2$,
indicating that $t\geq \tau_{\mathrm{BD}}$. With respect to the solution $f_{+}$,
we first consider a simple case that $r_2=r_7=0$, which means the diagonal
entries of the density matrix are all $1/3$. In this case, $\cos(\Delta\tau_{\mathrm{BD}})$
reduces to $\cos(\frac{\Theta}{2})$ and $f_{+}$ reduces to
\begin{equation}
f_{+}=\frac{-y+\sqrt{16x^2\cos^2\left(\frac{\Theta}{2}\right)+8xy\cos\Theta+y^2}}{4x}.
\end{equation}
A non-negative $\cos(\frac{\Theta}{2})-f_{+}$ means
\begin{equation*}
4x\cos\left(\frac{\Theta}{2}\right)+y\geq
\sqrt{16x^2\cos^2\left(\frac{\Theta}{2}\right)+8xy\cos\Theta+y^2}.
\end{equation*}
As both sides are positive, this inequality can be further simplified by taking
the square on both sides, which is of the form
\begin{equation}
8xy\left[\cos\left(\frac{\Theta}{2}\right)-\cos\Theta\right]\geq 0.
\end{equation}
This inequality naturally holds since $\Theta\in(0,\pi]$ and $x,y\geq 0$. Hence,
$t\geq\tau_{\mathrm{BD}}$ for the states with $r_2=r_7=0$. In the meantime, in
the regime $2x+y\leq 2|\vec{r}|^2\sin^2(\frac{\Theta}{2})$, $f_{+}$ is still
negative. Hence, in the crossover regimes between (\ref{eq:apx_regime1}),
(\ref{eq:apx_regime2}) and $2x+y\leq 2|\vec{r}|^2\sin^2(\frac{\Theta}{2})$,
$t$ is always not smaller than $\tau_{\mathrm{BD}}$.

All the three regimes discussed above are plotted in Fig.~\ref{fig:apx_3levelproof}(b)
as the red (dark gray) areas. In the regime $2x+y\geq
2|\vec{r}|^2\sin^2(\frac{\Theta}{2})$, the situation is a little complicated.
Now we first consider the case that $x$ and $y$ are fixed, which means $x+y$
is also fixed. Due to the fact $r^2_2+r^2_7=|\vec{r}|^2-x-y$, this condition
indicates $r^2_2+r^2_7=c$ is also fixed ($c$ a real constant). Then according
to the discussion in the proof of Theorem~\ref{lamma_threelevel}, the maximum
value of $\tau_{\mathrm{BD}}$ becomes
\begin{equation}
\max_{r_2,r_7}\,\tau_{\mathrm{BD}}=\frac{\Theta}{2\Delta\sqrt{1-\frac{4}{3}
(|\vec{r}|^2-x-y)}}:=\tau_{\mathrm{a},1}
\end{equation}
for $x+y\geq |\vec{r}|^2-1/3$ and
\begin{equation}
\max_{r_2,r_7}\,\tau_{\mathrm{BD}}\!=\!\frac{\Theta}{2\Delta\sqrt{1\!-\!\left[\frac{1}{2}
+\sqrt{\frac{1}{3}(|\vec{r}|^2\!-\!x\!-\!y\!-\!\frac{1}{4})}\right]^{2}}}
\!:=\!\tau_{\mathrm{a},2}
\end{equation}
for $x+y\leq |\vec{r}|^2-1/3$. Since the solutions of evolution time
$t$ (to reach the target $\Theta$) are related only to $x$ and $y$, $t$ is fixed
once $x$ and $y$ are fixed. If $t$ is indeed larger than $\tau_{\mathrm{a},1}$ and
$\tau_{\mathrm{a},2}$, then for all the states within the circle $r^2_2+r^2_7=c$,
$\tau_{\mathrm{BD}}$ is a valid lower bound. Otherwise, $\tau_{\mathrm{BD}}$
fails to be a lower bound at least for the states on the circle. To provide an
intuitive picture of this, we randomly generate 10000 states in the regime
$2x+y\geq 2|\vec{r}|^2\sin^2(\Theta/2)$ in the case of $|\vec{r}|^2=1$
and $\Theta=\pi/3$, as shown in Fig.~\ref{fig:apx_3levelproof}(b), to test if
$\tau_{\mathrm{a},1}$ and $\tau_{\mathrm{a},2}$ are lower than $t$. It can be
seen that though $\tau_{\mathrm{BD}}$ remains a valid lower bound for most
states (orange dots), the violation (blue dots) indeed happens. The borderline
is nothing but the equation
\begin{equation}
\cos(\Delta\tau_{\mathrm{a,i}})-f_{+}=0,
\end{equation}
where $i=1$ for $x+y\geq |\vec{r}|^2-1/3$ and $i=2$ for $x+y\leq |\vec{r}|^2-1/3$.
Substituting the expression of $f_{+}$ in Eq.~(\ref{eq:apx_cosdt}) into the equation
above, it reduces to
\begin{equation}
x\sin^2(\Delta \tau_{\mathrm{a},i})+y\sin^2\left(\frac{\Delta\tau_{\mathrm{a},i}}{2}\right)
=|\vec{r}|^2\sin^2\!\left(\frac{\Theta}{2}\right).
\end{equation}
Hence, in the regime
\begin{equation}
x\sin^2(\Delta \tau_{\mathrm{a},i})+y\sin^2\left(\frac{\Delta\tau_{\mathrm{a},i}}{2}\right)
\leq |\vec{r}|^2\sin^2\!\left(\frac{\Theta}{2}\right),
\end{equation}
the Bhatia-Davis formula is a valid lower bound.

Furthermore, due to the fact that the violation regime
\begin{equation}
x\sin^2(\Delta \tau_{\mathrm{a},i})+y\sin^2\left(\frac{\Delta\tau_{\mathrm{a},i}}{2}\right)
\geq |\vec{r}|^2\sin^2\!\left(\frac{\Theta}{2}\right)
\end{equation}
is within the regime $y\leq 4x$. Together with Eqs.~(\ref{eq:apx_regime1}) and
(\ref{eq:apx_regime2}), the full regime in which $\tau_{\mathrm{BD}}$ is a valid
lower bound is
\begin{equation}
\begin{cases}
y \geq 4|\vec{r}|\sin\left(\frac{\Theta}{2}\right)\sqrt{x}-4x, \\
y\geq 4x, \\
y\geq |\vec{r}|^2\sin^2\left(\frac{\Theta}{2}\right), \\
y\leq |\vec{r}|^2-x,
\end{cases}
\end{equation}
and
\begin{equation}
\begin{cases}
y \geq 4|\vec{r}|\sin\left(\frac{\Theta}{2}\right)\sqrt{x}-4x, \\
y\leq 4x, \\
y\leq |\vec{r}|^2-x, \\
y\sin^2(\frac{\Delta\tau_{\mathrm{a},i}}{2})\leq |\vec{r}|^2\sin^2\!(\frac{\Theta}{2})
-x\sin^2(\Delta \tau_{\mathrm{a},i}).
\end{cases}
\end{equation}
Theorem~\ref{theorem:3level} is then proved. \hfill $\blacksquare$

\end{document}